\documentclass[12pt]{iopart}

\expandafter\let\csname equation*\endcsname\relax
\expandafter\let\csname endequation*\endcsname\relax
\usepackage{graphicx}
\usepackage{lineno}
\usepackage{hyperref}
\hypersetup{colorlinks,linkcolor={blue},citecolor={blue},urlcolor={blue}}
\usepackage{tcolorbox}
\usepackage{gensymb}
\usepackage{siunitx}
\usepackage{amsmath}
\usepackage{amstext}
\usepackage{multicol}
\usepackage{subcaption}
\usepackage{indentfirst}
\usepackage{booktabs}
\begin{document}

%\title[Author guidelines for IOP Publishing journals in  \LaTeXe]{How to prepare and submit an article for 
%publication in an IOP Publishing journal using \LaTeXe}

\title[]{The effect of ion rotational flow on Hall thruster azimuthal instability via two dimensional PIC simulations}

\author{Zhijun Zhou\textsuperscript{1}, Lihuan Xie\textsuperscript{1}, Xin Luo\textsuperscript{2}, Yinjian Zhao\textsuperscript{1,*} and Daren Yu\textsuperscript{1}}
\ead{zhaoyinjian@hit.edu.cn}
\address{\textsuperscript{1}School of Energy Science and Engineering, Harbin Institute of Technology, Harbin 150001, People’s Republic of China}
\address{\textsuperscript{2}College of Aerospace and Civil Engineering, Harbin Engineering University, Harbin 150001, People’s Republic of China}

\vspace{10pt}
\begin{indented}
\item[]\today
\end{indented}

\begin{abstract}
Previous experimental studies have found that the neutral gas rotational flow in the opposite direction of electron Hall drift can lead to better
experimental results comparing to the same direction. In Hall thrusters, the core factor influencing
operational states is the electron cross field transport, where the azimuthal instability serves as a key mechanism. The rotational flow of neutral gas may affect instability by altering initial azimuthal 
velocity of ions, which has not been investigated before.
Therefore, to study the effects of ion rotational
flow of varying magnitudes and directions on azimuthal instability, simulations are conducted in this work
based on two
benchmark particle-in-cell (PIC) cases: the azimuthal-axial and the azimuthal-radial. 
The results indicate that the ion
rotational flow velocity can potentially complicate the coupling characteristics of the electron cyclotron drifting instability and the modified two stream instability, particularly when a reverse rotational flow velocity is added.
In general, both co-directional and reverse ion
rotational flow have been observed to inhibit azimuthal instability,
which results in a decrease in axial electron mobility.
A 1\% addition of the ion rotational flow (compared to the electron drift) would result in a 10\% change of the electron mobility due to varied azimuthal instability, and the decrease in electron mobility of the reverse ion rotational flow is
greater than that of co-directional.
In addition, detailed spectral analyses are carried out to study the relation between ECDI, MTSI, and resonant wave-wave interactions.
\end{abstract}

\section{Introduction}

Hall thruster (HT) is a space electric propulsion device with high technology maturity.
It has the advantages of simple structure, high specific impulse, long lifetime, and is widely used in space missions such as spacecraft attitude control, orbital maneuvering, and position holding\cite{9697981}. 
At present, the number of satellites launched by Starlink low-orbit satellite constellation has reached 7000, and these satellites are all using HTs in orbit. 
Such a huge market demand has prompted researchers all over the world to explore the performance optimization measures of HT.

In order to improve the propellant ionization rate of HT, Zhang et al.\cite{Xu_2017} first proposed a design scheme of rotational propellant injection (RPI). 
By adding an azimuthal diversion rail behind the gas distributor, the neutral atom had a certain azimuthal velocity component, thus the axial velocity was reduced, the atom density was increased in the discharge channel as well as the HT performance. 
Due to the significant increase in the propellant ionization rate of RPI, Xia et al.\cite{WOS:000524257600025,WOS:000580600700043} designed the gas distributor outlet holes as inclined holes and applied it to the HT using krypton propellant, which alleviated the problem that krypton was difficult to be ionized.
Similar to the RPI, the vortex injection method converts the neutral gas flow from axial to rotational. 
Satpathy et al.\cite{Satpathy_2024} designed outlet holes tangentially along the circumference of the gas distributor and applied vortex injection mode to the Thrusters with Anode Layers (TAL), increasing the ionization rate of the argon propellant from 26.5\% to 30.6\%. 
Ding et al.\cite{10.1063/1.4986007} applied the vortex injection mode to the low-power cylindrical Hall thruster (CHT), which significantly improved the ionization intensity of the main ionization region near the anode. 
In general, increasing the rotational flow velocity of the propellant can significantly increase the propellant ionization rate and can be widely used in multi-propellant and multi-type HTs.

In addition, we also notice that Ding et al.\cite{10.1116/1.5037215} experimentally studied the matching relationship between RPI and the direction of the radial magnetic field. The angle between the azimuthal diversion rail and the axis of the HT was $\beta$=\ang{70}. While keeping the angle unchanged, the direction of electron drifting velocity was reversed by reversing the direction of the radial magnetic field. 
When the direction of the electron drift was changed from the same direction as the rotational velocity of the RPI to the opposite direction, the anode efficiency and the propellant utilization rate of the HT were increased by 1.78\% and 3\% on average. 
The explanation of this phenomenon in Ref.\cite{10.1116/1.5037215} was that the electron drift in the opposite direction of RPI increased the electron-neutral collision frequency, and therefore improved the performance of the HT.

If we examine the influence of RPI on the electron-neutral collision frequency, which is a function of the neutral atomic
number density $n_n$, the temperature of the electron $T_e$, and the velocity of electrons relative to neutral atoms\cite{frias2016plasma}
\begin{equation}\label{eq:nu_m_c}
\nu_{en}=\sigma_{en}n_{n}(v_e- {V}_{arot}),
\end{equation}
where $\sigma_{en}$ is the effective electron-neutral scattering cross-section, and it is related to
the electron temperature.
The rotational propellant injection brings two effects, one is that the $n_n$ is increased by about 20$\%$\cite{Xu_2017} due to the decrease of axial velocity, and the other is that the atoms have a certain initial azimuthal velocity, which leads to azimuthal effects. 
Thus the atomic number density is essentially the same for the opposite and the co-directional rotational flow because of the same axial velocity. 
For typical parameters in HT, $n_n =$ 2$\times$$10^{19}~\mathrm{m^{-3}}$, $T_e = 10$ eV, and the magnetic field in the main ionization zone $B_0$ = 100 Gauss, $v_e=\sqrt{8T_e/\pi m_e} \approx7.5$$\times$$10^5$ m/s.
According to the simulation results of SPT-100
neutral atom velocity by Boccelli et al.\cite{WOS:000727811800010}, using Direct Simulation Monte Carlo
method, when the anode temperature is 1000K, the neutral atom velocity would
not exceed $v_a$ = 400m/s, meaning the maximum rotational velocity of the RPI is only
about $V_{arot} = v_a\mathrm{cos}\beta \approx$ 376 m/s.
Therefore, the increase in electron-neutral collision frequency of the reverse RPI relative to the co-directional RPI is only 0.11\%,
which is one order of magnitude
less than
1.78\% and 3\% found in \cite{10.1116/1.5037215}.
% Obviously it is not suffcient to explain the
% related phenomenon from the perspective of the electron-neutral collision ionization and
Therefore, we aim to investigate this problem from the perspective of azimuthal instability, as this instability is expected to be sensitive to the ion rotational flow.

We generally consider that the initial velocity of ions produced by the ionization of atoms inherit the velocity of the atoms.
Thus each ion has a certain initial rotational flow velocity for RPI.
Notably, the electrons produced by the ionization are rapidly magnetized, thus we do not consider the effect of RPI on the azimuthal velocity of electrons.
The electrons in HT are magnetized while the ions are not. 
The huge \textbf{E}$\times$\textbf{B} drifting velocity of electrons relative to ions drives the microscopic plasma instabilities in HT, mainly including ECDI and MTSI\cite{10.1063/5.0145536}. 
The frequency of the first ECDI harmonic (1st ECDI) is approximately 5 MHz, the wavelength is about 1mm, and the direction is along \textbf{E}$\times$\textbf{B}.
The frequency of MTSI is about 1 MHz, the wavelength is several millimeters, and has a component along the direction of \textbf{B}.
Coupling development of microscopic azimuthal instabilities over time results in the anomalous transport of electrons across magnetic field lines.
Therefore, the ion
rotational flow velocity would affect the azimuthal instability of HT, 
and the anomalous electron transport.

The azimuthal instability of HT is the core factor affecting the anomalous electron transport\cite{10.1063/5.0010135}.
The research methods on azimuthal instability mainly include theoretical analysis, numerical simulation. 
The plasma dispersion equation in the linear development stage of the plasma azimuthal instability was derived mainly by linearization theory\cite{10.1063/1.2359718,10.1063/5.0122293,10.1063/1.4948496,Lafleur_2017}.
Lafler et al.\cite{Lafleur_2018,10.1063/1.5017033} simplified the dispersion relation for 1st ECDI in the framework of kinetic theory and gave an approximate expression for the growth rate.
However, the theoretical analysis contains a number of assumptions and cannot describe the nonlinear evolution of the plasma in HT.
Because the azimuthal instability in HT involves complex wave-particle interaction and has strong nonlinear characteristics, the current research on azimuthal instability is mainly carried out by 2D PIC simulations\cite{10.1063/5.0138223,10.1063/5.0176581,10.1063/1.5139035}. 
Croes et al.\cite{Croes_2017} investigated the evolution characteristics of ECDI by 2D radial-azimuthal simulations and demonstrated that 1st ECDI led to a strong increase in electron-ion friction.
Janhunen et al.\cite{10.1063/1.5033896} firstly proposed the MTSI found in 2D radial-azimuthal simulations played an important role in electron anomalous transport and electron radial heating.
Petronio et al.\cite{10.1063/5.0046843} found the formation condition of MTSI was related to |B| and |E|.

In summary, there is currently no study on the influence of ion
rotational flow on the azimuthal instability of HT. 
Therefore, 2D azimuthal-axial and azimuthal-radial PIC simulations are carried out in this paper to explore the influence mechanism of 
ion rotational flow velocity on the azimuthal instability and anomalous electron transport.
In Sec.~\ref{sec:theory}, according to the plasma dispersion relation, the variation between the instability and ion rotational flow velocity is analyzed theoretically.
In Sec.~\ref{sec:simulation model},
the basic parameter settings of 2D axial-azimuthal and radial-azimuthal PIC simulation model are described in detail.
The results of axial-azimuthal simulation and radial-azimuthal simulation are analyzed separately in Sec.~\ref{sec:azimuthal-axial simulation results} and Sec.~\ref{sec:azimuthal-radial simulation results}.
The discussions and conclusions are given in Sec.~\ref{sec:conclusion}.
In addition, the simulation tool applied in this work is the WarpX open-source PIC code\cite{warpx}. 
WarpX is a highly-parallel and highly-optimized code, which can run on GPUs and multi-core
CPUs.
WarpX has been used to simulate low-temperature plasmas in HT\cite{Xie_2024,Marks_2024}.

\section{Theoretical analysis\label{sec:theory}}

Ducrocq et al.\cite{10.1063/1.2359718} derived the dispersion relation that describes the high-frequency electron drift in HT in 2006.
The ion density perturbation was obtained by linearizing the ion cold fluid equation,
the electron density perturbation was obtained by linearizing the Vlasov equation of the electron following the Maxwellian distribution. 
The perturbation terms were then brought into the Poisson's equation, thus the general form of the oscillatory dispersion relation of plasma in HT was obtained. 
Cavalier et al.\cite{10.1063/1.4817743} then rewrote the dispersion equation by introducing the 
Gordeev function as
\begin{equation}\label{eq:dispersion equation}
1+k^{2}\lambda_{D}^{2}+\mathrm{g}\left(\frac{\omega-k_{y}V_{d}}{\omega_{ce}},(k_{x}^{2}+k_{y}^{2})\rho^{2},k_{z}^{2}\rho^{2}\right)-\frac{k^2\lambda_D^2\omega_{pi}^2}{\left(\omega-k_xv_p\right)^2}=0,
\end{equation}
where $g\left(\Omega,X,Y\right)$ is the Gordeev function, defined as follows 
\begin{equation}
\begin{aligned}
\mathrm{g}(\Omega,X,Y)& =i\Omega\int_{0}^{+\infty}e^{-X[1-\cos(\varphi)]-\frac{1}{2}Y\varphi^{2}+i\Omega\varphi}d\varphi \\
&=\frac{\Omega}{\sqrt{2Y}}\mathrm{e}^{-X}\sum_{m=-\infty}^{+\infty}Z\biggl(\frac{\Omega-m}{\sqrt{2Y}}\biggr)\mathrm{I}_{m}(X),
\end{aligned}
\end{equation}
where $Z(z)$ is the plasma dispersion function, $I_m(X)$ is the modified Bessel function of the first kind, $\omega_{pi}$ is ion plasma frequency, $\omega_{ce}$ is the electron cyclotron frequency, $v_p$ is axial velocity of ions. $k_x$, $k_y$, $k_z$ are the three orthogonal components of wave vector \textit{\textbf{k}}, with $k_x$ parallel to static electric field $\boldsymbol{E}=E_0\hat{\boldsymbol{x}}$, $k_z$ parallel to static magnetic field $\boldsymbol{B}=B_0\hat{\boldsymbol{z}}$, $\lambda_D$ is the
Debye length, $\rho=v_{the}/\omega_{ce}$ is the electron Larmor
radius at the thermal velocity $v_{the}=\sqrt{k_{B} T_{e} / m_{e}}$, $\boldsymbol{V_d}=-E_0/B_0\hat{\boldsymbol{y}}$ is the electron drifting velocity. Thus we add ion rotational flow velocity term $-k_yV_{rot}$ to the formula, the Eq.~(\ref{eq:dispersion equation}) becomes 
\begin{equation}\label{eq:M dispersion equation}
1+k^{2}\lambda_{D}^{2}+\mathrm{g}\left(\frac{\omega-k_{y}V_{d}-k_yV_{rot}}{\omega_{ce}},(k_{x}^{2}+k_{y}^{2})\rho^{2},k_{z}^{2}\rho^{2}\right)-\frac{k^2\lambda_D^2\omega_{pi}^2}{\left(\omega-k_xv_p-k_yV_{rot}\right)^2}=0 .
\end{equation}
Then Cavalier normalize the equation with $\hat{\omega}=\omega/\omega_{pi}$, $\hat{k}=k\lambda_{D} $, $\hat{V}=V/c_s$, $\mathrm{\hat{M}=M/m_e}$, $c_s=\lambda_{D}\omega_{pi}$. The normalized quantity with ion rotational flow velocity term reads
\begin{equation}\label{eq:M normalized equation}
\left(\hat{\omega}-\hat{k}_{x}\hat{v}_{p}-\hat{k}_{y}\hat{V}_{rot}\right)^{2}=\frac{\hat{k}^{2}}{1+\hat{k}^{2}+\mathrm{g}\left(\frac{\hat{\omega}-\hat{k}_{y}\hat{V}_{d}-\hat{k}_{y}\hat{V}_{rot}}{\hat{\omega}_{ce}},\left(\hat{k}_{x}^{2}+\hat{k}_{y}^{2}\right)\frac{\hat{M}}{\hat{\omega}_{ce}^{2}},\hat{k}_{z}^{2}\frac{\hat{M}}{\hat{\omega}_{ce}^{2}}\right)}.
\end{equation}
Finally the iterative equation becomes
\begin{equation}\label{eq:iterative equation}
\left.\left\{\begin{array}{l}\hat{\omega}_{+,n+1}=\hat{k}_{x}\hat{v}_{p}+\hat{k}_{y}\hat{V}_{rot}+\hat{\omega}_{r,n+1}+i\epsilon\hat{\gamma}_{n+1}\\\hat{\omega}_{-,n+1}=\hat{k}_{x}\hat{v}_{p}+\hat{k}_{y}\hat{V}_{rot}-\hat{\omega}_{r,n+1}-i\epsilon\hat{\gamma}_{n+1}\end{array}\right.\right.,
\end{equation}
with $\hat{\omega}_{r,n+1}$ and $\hat{\gamma}_{n+1}$ as follows
\begin{equation}
\begin{cases}\widehat{\omega}_{r,n+1}=\frac{1}{\sqrt{2}}\frac{\widehat{k}}{\sqrt{h_{n}^{2}+g_{in}^{2}}}\biggl(h_{n}+\sqrt{h_{n}^{2}+\mathrm{g}_{in}^{2}}\biggr)^{\frac{1}{2}}\\\\\widehat{\gamma}_{n+1}=\frac{1}{\sqrt{2}}\frac{\widehat{k}}{\sqrt{h_{n}^{2}+g_{in}^{2}}}\biggl(h_{n}+\sqrt{-h_{n}^{2}+\mathrm{g}_{in}^{2}}\biggr)^{\frac{1}{2}}\end{cases},
\end{equation}
where $\epsilon$ is the sign
of $-\mathrm{g}_{in}$, $h_n=1+\hat{k}^2+\mathrm{g}_{r n}$, $\mathrm{g}_{rn}$ and $\mathrm{g}_{in}$ are the real and imaginary part of the Gordeev function respectively.

\begin{figure}[htbp]
    \centering
    \begin{minipage}[h]{0.35\textwidth}
    \makebox[1em]{a)}
    \centerline{\includegraphics[scale=0.25]{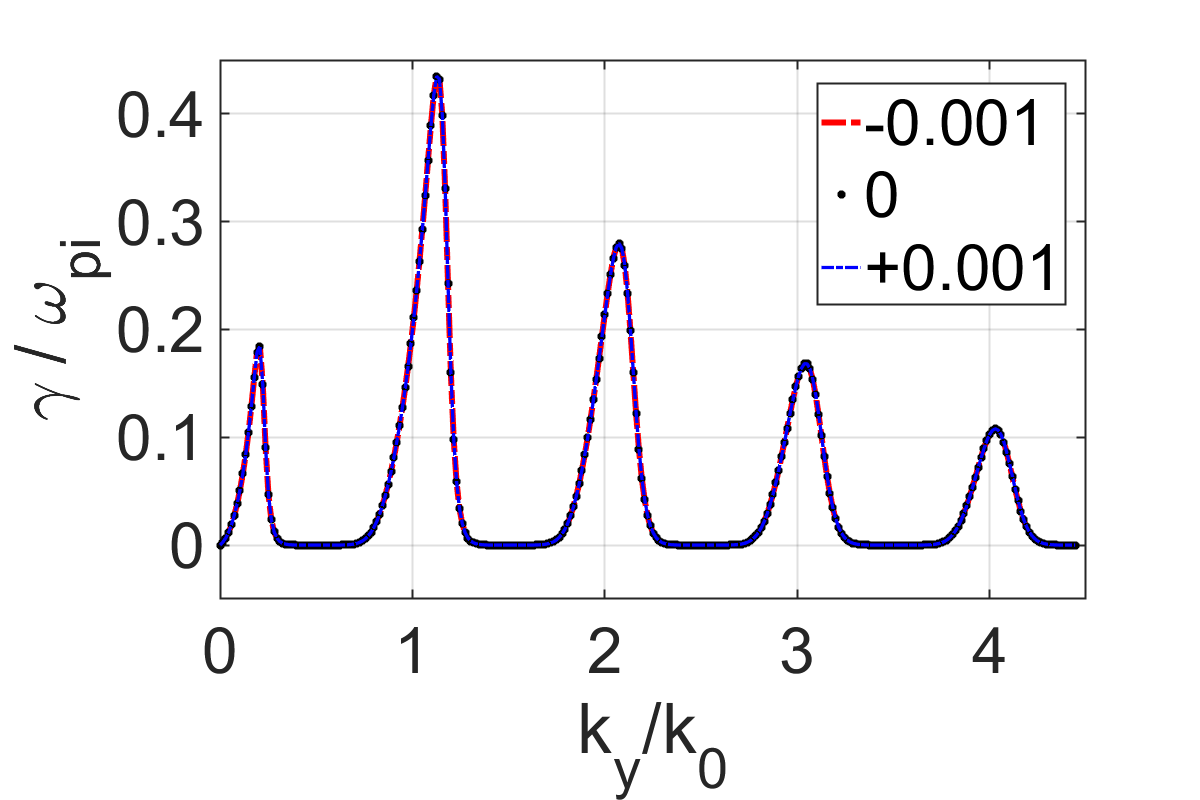}}
    \makebox[1em]{c)}
    \centerline{\includegraphics[scale=0.25]{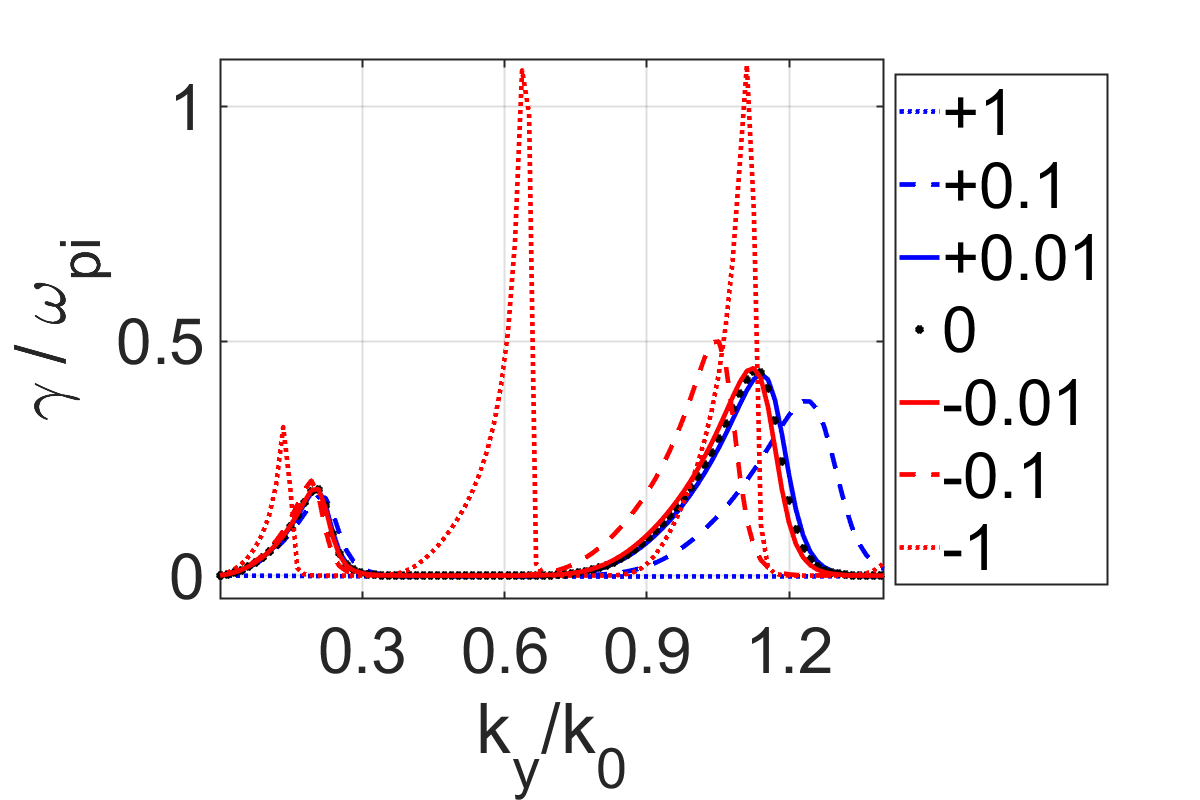}}
    \end{minipage}
    \hspace{50pt}
    \centering
    \begin{minipage}[h]{0.35\textwidth}
    \makebox[1em]{b)}
    \centerline{\includegraphics[scale=0.25]{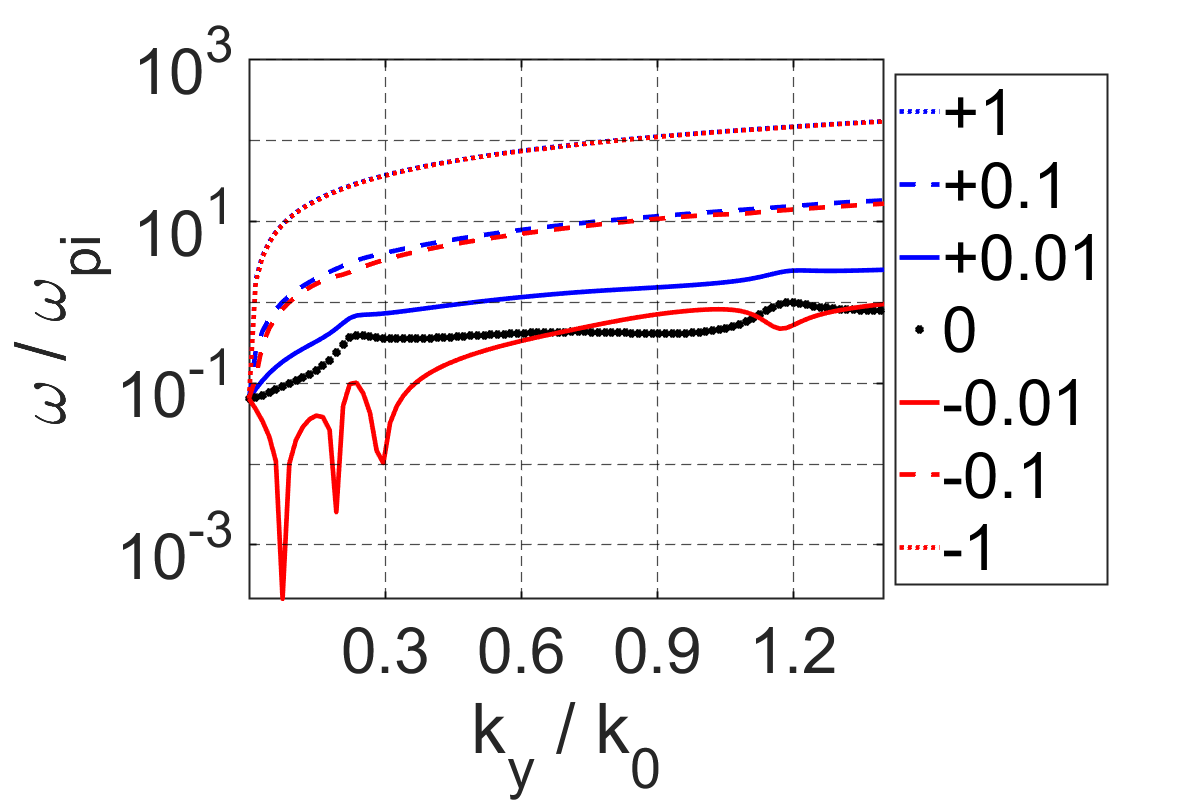}}
    \makebox[1em]{d)}
    \centerline{\includegraphics[scale=0.25]{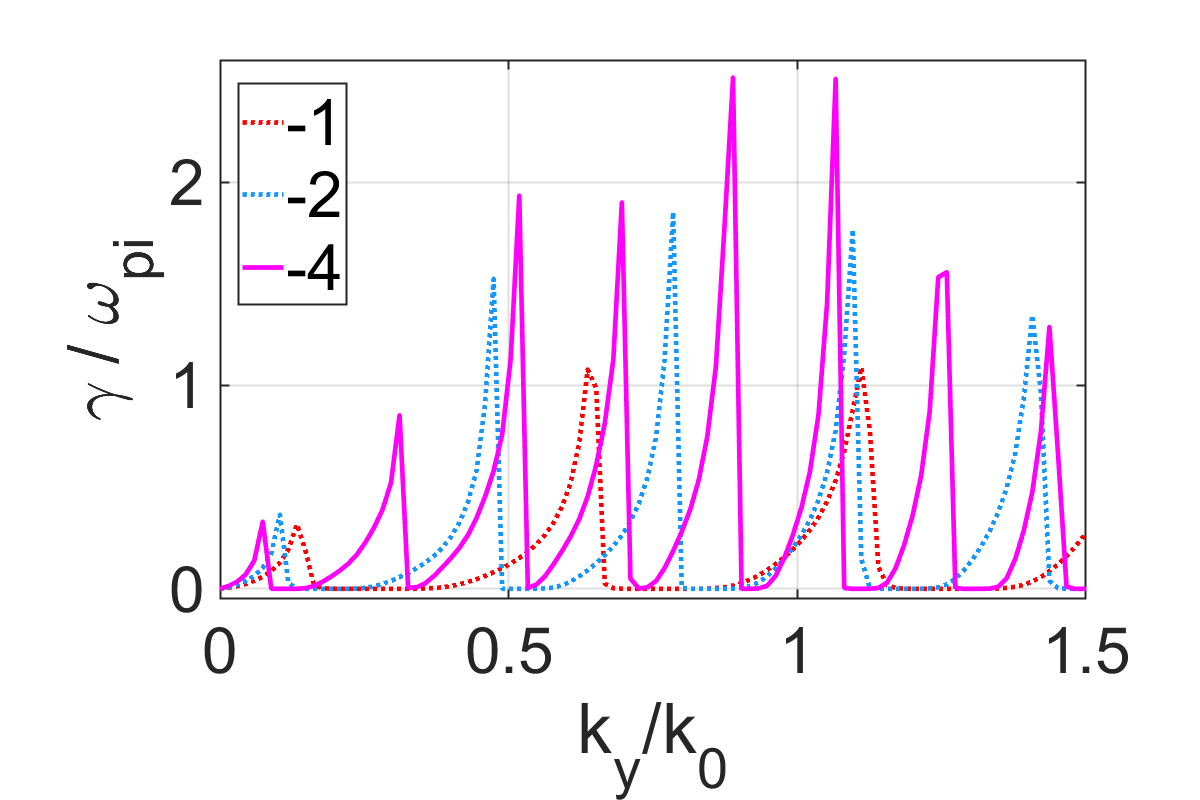}}
    \end{minipage}
\caption{Iterative results of Eq.\ref{eq:iterative equation} at different ion
rotational flow velocities, change of a) growth rates for $\alpha$ = 0, $\pm$0.001, b) frequencies after taking absolute values, c) growth rates for $\alpha$ = 0, $\pm$0.01, $\pm$0.1, $\pm$1, d) growth rates for $\alpha$ = -1, -2, -4 with azimuthal wavenumber.}
\label{fig:theory}
\end{figure}

Based on the iterative format obtained above, dispersion relation with different magnitude and direction of ion rotational flow velocity can be solved numerically.
On the basis of electron drifting velocity, here we define $\alpha=\boldsymbol{(V_{rot}\cdot V_d)}/{V_d}^2$ as a measure of magnitude and direction.
Subsequently, we performed numerical iterative solutions for servel cases, and the results are shown in Fig.~\ref{fig:theory}, the normalized frequencies in  Fig.~\ref{fig:theory} \textcolor{blue}{b)} have been obtained by taking absolute values.
Here $k$ is normalized by $k_0 = \omega_{ce} / V_{d}$, $n = 6\times 10 ^{16}~\mathrm{m^{-3}}$, $T_e$ = 10 eV, $B_0$ = 200 Gauss, $E_0$ = $1\times10^4~\mathrm{V/m}$, $\hat{v_p}$ = 3.16, axial wavenumbrer component $\hat{k_x}$ = 0.012, radial wavenumber component $\hat{k_z}$ = 0.0236, the relative error is set to $10^{-13}$.

It is clear that the introduction of an ion rotational flow velocity gives rise to a Doppler shift in the frequency, with $\omega \rightarrow \omega-k_yV_{rot}$. Here $V_{rot}$ = -5$\times$$10^5$$\alpha~\mathrm{m/s}$ in the coordinate system employed in the theoretical derivation, so the magnitude of frequencies under the co-directional ion rotational flow is greater than that under the reverse ion rotational flow at the same velocity.
Below we use the change of the frequency at $k_y=k_0$ to illustrate the effect of adding the ion rotational flow.
When $\alpha$ = 0, $\pm$0.001, the lines of growth rate almost coincide as $k_0V_{rot}\ll \omega_{\alpha=0}$, as
illustrated in Fig.~\ref{fig:theory} \textcolor{blue}{a)}.
For cases $\alpha$ = $\pm$0.01, $k_0V_{rot}\sim \omega_{\alpha=0}$, the evident frequency peaks of MTSI and 1st ECDI, are clearly discernible as
illustrated in Fig.~\ref{fig:theory} \textcolor{blue}{b)}.
When $\alpha$ = -0.01, the overall reduction in frequencies results in $\omega < 0$ when $k_y/k_0 \in$ [0.074, 0.19] and [0.29, +$\infty$]. 
The symbol of $\omega$ indicates the direction in which a wave propagates, and there are simultaneously unstable waves that propagate in the opposite direction for case $\alpha$ = -0.01.
When $\alpha$ = $\pm$0.1, $\pm$1, 
since $k_0V_{rot}\gg \omega_{\alpha=0}$, the frequency is linearly related to the wavenumber, and the frequency peaks of the MTSI and 1st ECDI are masked by $k_0V_{rot}$.

\begin{figure}
	\centering
	\includegraphics[width=0.5\linewidth]{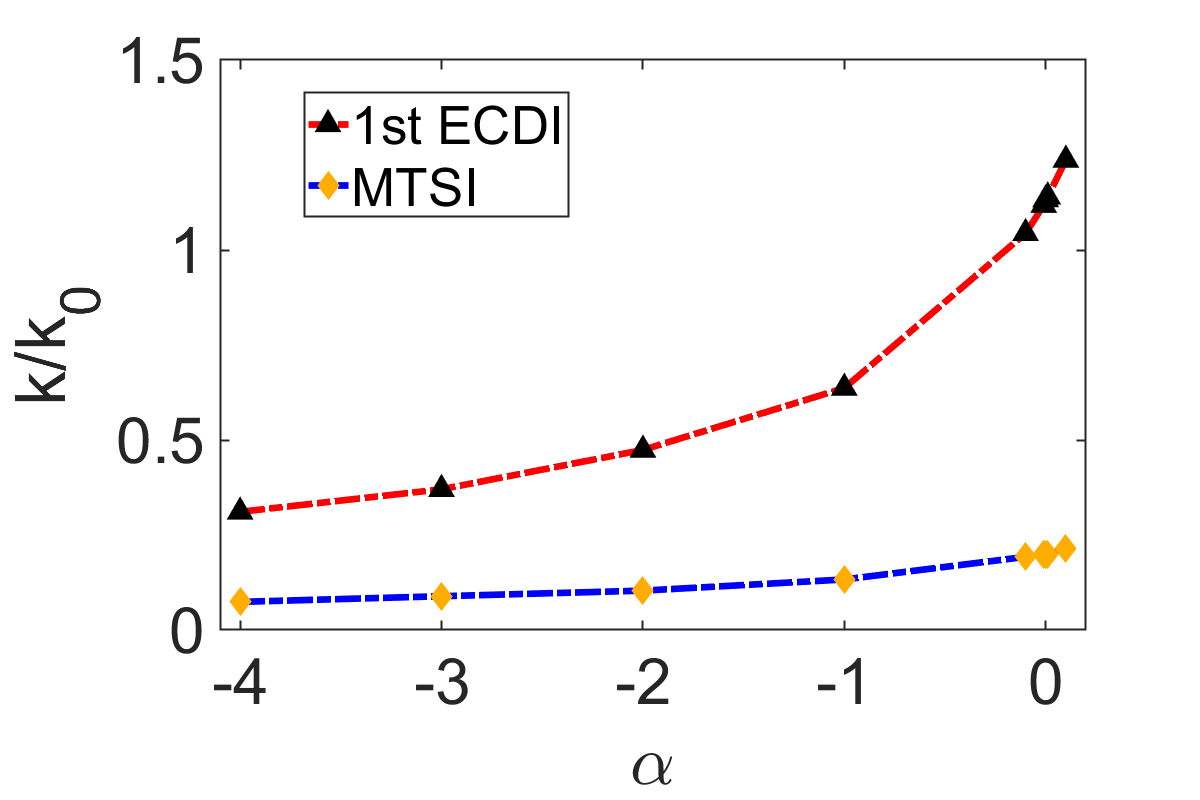}
	\caption{Wavenumbers of MTSI and the 1st ECDI in different ion rotational flow
velocities}
	\label{fig:wavenumber}
\end{figure}

Then we analyze the effect of the ion rotational flow on growth rates.
In the condition of $-1 \leq \alpha \leq1$, as the relative velocity of electrons and ions decreases, 
the growth rates of MTSI and the 1st ECDI also decrease.
Particularly, when $\alpha$ = 1, there is no relative drift between electrons and ions, thus the instability can not be triggered and the growth rate is zero, as shown in Fig.~\ref{fig:theory} \textcolor{blue}{c)}. 
We also find that the wavenumbers of the 1st ECDI and MTSI decrease as the velocity of ions relative to electrons increase, as shown in Fig.~\ref{fig:theory} \textcolor{blue}{c), d)}.
Fig.~\ref{fig:wavenumber} shows the decrease of wavenumbers for MTSI and the 1st ECDI.

In general, the theoretical results demonstrate a number of intriguing conclusions, 
particularly when adding reverse ion
rotational flow. Given that the impact of $|\alpha| = 0.001$ on instability is minimal compared to $\alpha$ = 0, it is decided not to undertake further simulations in this regard. 
And it is straightforward to incorporate the ion rotational flow velocity into the simulation setup, so we conduct simulations in which more extreme ion rotational flow velocities ($\alpha$ = $\pm$1) are introduced.

In the end, we have selected seven cases with $\alpha$ = 0, $\pm$0.01 ($\sim 10^3~\mathrm{m/s}$), $\pm$0.1 ($\sim 10^4~\mathrm{m/s}$), $\pm$1 ($\sim 10^5~\mathrm{m/s}$) for our subsequent simulations, 
but we remind readers that the ion rotational speed in practice is usually a few hundred meters per second. 
Therefore, in the following simulations, those cases $|\alpha| \geq 0.1$ are deliberately chosen beyond their practical ranges to explore hypothetical scenarios and assess the system's theoretical response.
\section{Two dimensional simulation setup\label{sec:simulation model}}
\subsection{Two dimensional axial-azimuthal simulation setup} 
\begin{figure}[ht]
\centering
    \begin{minipage}[h]{0.35\textwidth}
    \makebox[1em]{a)}
    \centerline{\includegraphics[scale=0.35]{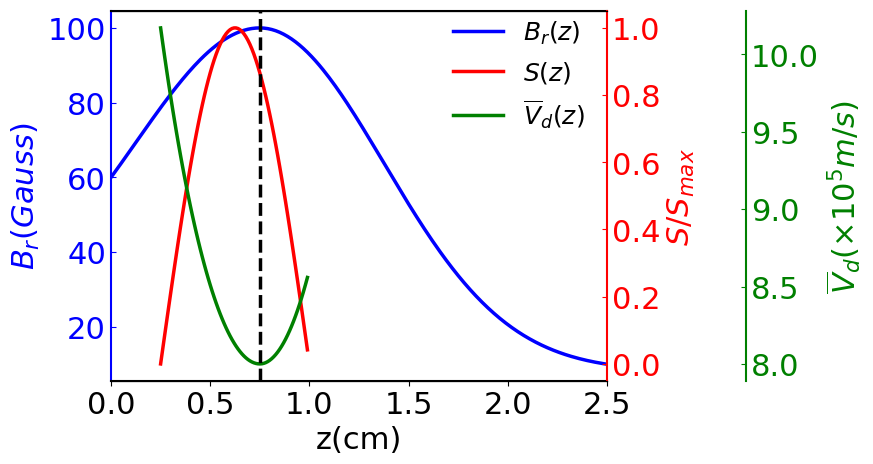}}
    \end{minipage}
    \hspace{30pt}
    \centering
    \begin{minipage}[h]{0.35\textwidth}
    \makebox[1em]{b)}
    \centerline{\includegraphics[scale=0.35]{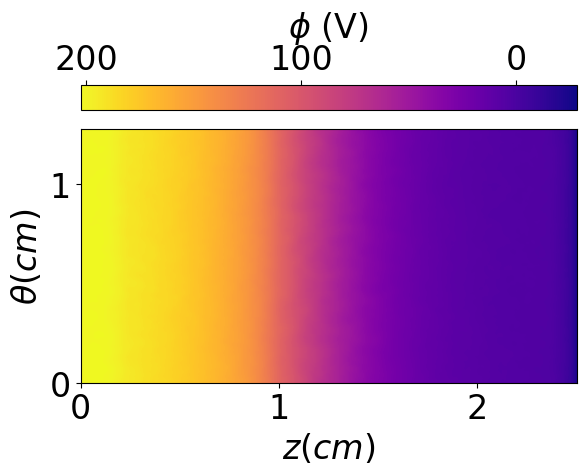}}
    \end{minipage}
\caption{a) Axial profiles of radial magnetic field intensity $B_r$, normalized ionization rate S and mean electron drifting velocity $\overline{V}_d$, the black dashed line indicates the position of maximum magnetic field, b) electric potential distribution in the 2D axial-azimuthal benchmark at t = 0.25 $\mu$s.}
\label{fig:BSVd}
\end{figure}

The two dimensional axial-azimuthal simulation setup is mainly based on the benchmark given by Charoy et al.\cite{Charoy_2019}. 
We set the z direction to be the axial direction, the y direction to be the azimuthal direction, and the azimuthal curvature is not considered. 
The axial length is $L_z$ = 2.5 cm 
with 512 evenly distributed grids, the azimuthal size is $L_y$ = 1.28 cm with 256 evenly distributed grids. 
The initial plasma is evenly distributed throughout the simulation domain with density $n_0=5\times10^{16}~\mathrm{m^{-3}}$, the initial electron temperature is $T_e$ = 10 eV with isotropic Maxwellian velocity distribution, the initial ion temperature is $T_i$ = 0.5 eV with isotropic Maxwellian velocity distribution. 
Thus the Debye length is $ \lambda_{D}=\left(\varepsilon_{0} k_{B} T_{e 0} / n_{0} e^{2}\right)^{1 / 2} \approx 0.105 \mathrm{~mm} $,
both axial and azimuthal individual grid sizes are smaller than the Debye length. 
Set the simulation time step  to $\Delta t= 5\times10^{-12}$ s and the total simulation time to $20\mathrm{~\mu s}$. The number of macroparticles per cell is $N_{ppc}$ = 70. 
The Dirichlet boundary condition with constant potential $\phi_{z=0}$ = 200 V is used in left side of simulation domain.
Periodic boundary conditions are applied to the upper and lower boundaries of the simulation domain.
For the cathode boundary, we set the electron injection position and the cathode potential 
according to the benchmark. 
The potential at the cathode electron injection position is $\phi_{z=z_{in}}$ = 0 V and $z_{in}$ = 2.4 cm.
When a particle leaves the computational domain from the left or right boundaries, the particle is deregistered.
The radial magnetic field has only an axial gradient and the distribution of the radial magnetic field along the axis is Gaussian, as follows
\begin{equation}
B_{x}(z)=a_{k}\exp\left(-\frac{(z-z_{B_{max}})^{2}}{2\sigma_{k}^{2}}\right)+b_{k},
\end{equation}
where the axial position of the maximum magnetic field is $z_{Bmax}$ = 0.75 cm. 
Taking the maximum magnetic field strength as the demarcation point, the axial distribution functions of the magnetic field are Gaussian distributions of different shapes, as illustrated 
the blue line in Fig.~\ref{fig:BSVd} \textcolor{blue}{a)}.
When $z<z_{B_{max}}$, k = 1, $B_0=B(z=0)$ = 60 Gauss. When $z>z_{B_{max}}$, k = 2, $B_{L_z}=B(z=L_z)$ = 10 Gauss.
Taking $\sigma_{1} = \sigma_{2} =0.25 L_z$ = 0.625 cm, we get $a_1$ = 77.94 Gauss, $b_1$ = 22.06 Gauss,
$a_2$ = 91.82 Gauss, $b_2$ = 8.18 Gauss.

Regardless of the effect of the ionization process, the ionization rate distribution function is used to generate plasma each step in the benchmark, as follows 
\begin{equation}
\begin{cases}S(z)=S_0\cos(\pi\frac{z-z_m}{\mathrm{z_2}-\mathrm{z_1}}) &\mathrm{z}\in(\mathrm{z_1},\mathrm{z_2})\\\\S(z)=0&\text{else}\end{cases},
\end{equation}
where $z_1$ = 0.25 cm, $z_2$ = 1 cm, $z_{m}=(z_{1}+z_{2})/2=0.625 ~\mathrm{cm}$. 
By setting the current density to 400 $\mathrm{A/m^2}$, we obtained the value of the ionization source term $S_0=5.23\times10^{23}~\mathrm{m^{-3}s^{-1}}$ after integration. 
The normalized ionization rate distribution along the axial direction is shown by the red line in Fig.~\ref{fig:BSVd} \textcolor{blue}{a)}.
For the initial arrangement and subsequent additions of ions in the simulation domain, we set the ion rotational flow velocity with the following distribution
\begin{equation}
V_{rot}(z)= \alpha \overline{V}_d=\frac{\alpha(\phi _{z=0}-\phi _{z=z_{in}})}
{z_{in}B(z)}
\end{equation}
As shown in Fig.~\ref{fig:BSVd} \textcolor{blue}{b)}, at 0.25 $\mu$s, the axial potential is almost linear, so we use $(\phi _{z=0}-\phi _{z=z_{in}})/z_{in}$ to represent the initial axial electric field in the simulation domain. In this way we estimate roughly the mean electron drifting velocity $\overline{V}_d$, as illustrated 
the green line in Fig.~\ref{fig:BSVd} \textcolor{blue}{a)}.
Seven cases are simulated using WarpX under the condition of $\alpha$=0, $\pm1$, $\pm0.1$, $\pm0.01$. Parallel computing with 64 MPI and 4 OpenMP takes 11.5 days for a single case.

\subsection{Two dimensional radial-azimuthal simulation setup} 
The two dimensional radial-azimuthal simulation setup is mainly based on the benchmark given by Villafana et al.\cite{Villafana_2021}. 
We set the x direction to be the radial direction, the y direction to be the azimuthal direction, and azimuthal curvature is also not considered. 
The axial length and azimuthal length are both set to $L_x=L_y$ = 1.28 cm with 256 evenly distributed grids. 
The initial plasma is evenly distributed throughout the simulation domain with density $n_0=5\times10^{16}\mathrm{~m^{-3}}$, the initial electron temperature is $T_e$ = 10 eV with isotropic Maxwellian velocity distribution, the initial ion temperature is $T_i$ = 0.5 eV with isotropic Maxwellian velocity distribution. 
Set the simulation time step  to $\Delta t= 1.5\times10^{-11}$ s and total simulation time to $30\mathrm{~\mu s}$. The number of macroparticles per cell is $N_{ppc}$ = 100.
The Dirichlet boundary condition with constant potential $\phi$ = 0 V is used for both the inner and outer walls in the radial direction, and periodic boundary conditions are used in the azimuthal direction.
The radial magnetic field intensity is set to $B_0$ = 200 Gauss.
Virtual axial boundaries with a thickness of 1 cm is added for injecting and deregistering high-speed particles.
Axial electric field is set to $E_0 = 1.0\times10^4$ V/m.
Therefor, ion rotational flow velocity is $V_{rot}=\alpha {E_0}/{B_0} = 5 \times 10^5\alpha$ m/s. 
Similarly seven cases are simulated using WarpX under the condition of $\alpha$ = 0, $\pm1$, $\pm0.1$, $\pm0.01$. And parallel computing with 256 MPI and 1 OpenMP takes 32 hours for a single case.

\section{Result analyses of axial-azimuthal simulations\label{sec:azimuthal-axial simulation results}}
\subsection{Effects on plasma profiles}

\begin{figure}[htbp]
    \centering
    \begin{minipage}[h]{0.35\textwidth}
    \makebox[1em]{a)}
    \centerline{\includegraphics[scale=0.35]{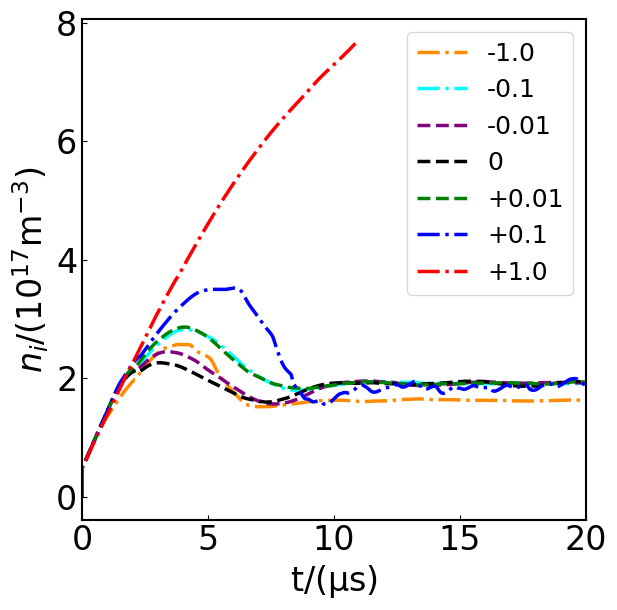}}
    \end{minipage}
    \hspace{17pt}
    \centering
    \begin{minipage}[h]{0.35\textwidth}
    \makebox[1em]{b)}
    \centerline{\includegraphics[scale=0.35]{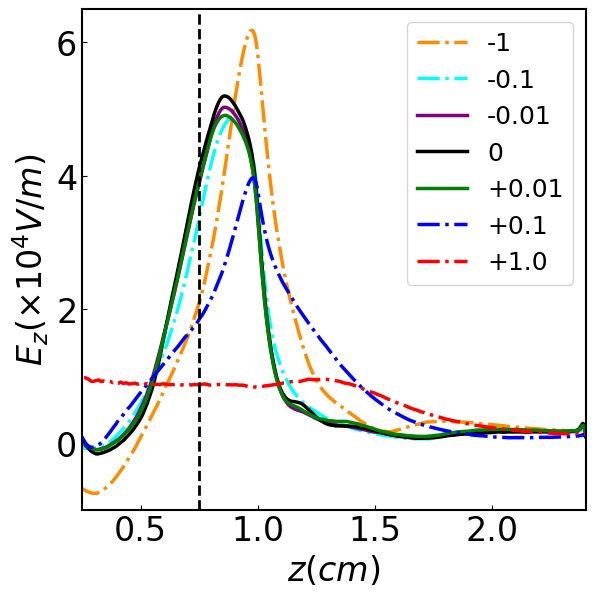}}
    \end{minipage}
\caption{a) Time profiles of ion density, b) axial profiles of the mean axial electric field over 10-20 $\mu$s, the black dashed line indicates
the position of maximum magnetic field.}
\label{fig:AZ_ni_t}
\end{figure}
\begin{figure}[ht]
\centering
\includegraphics[width=0.9\textwidth]{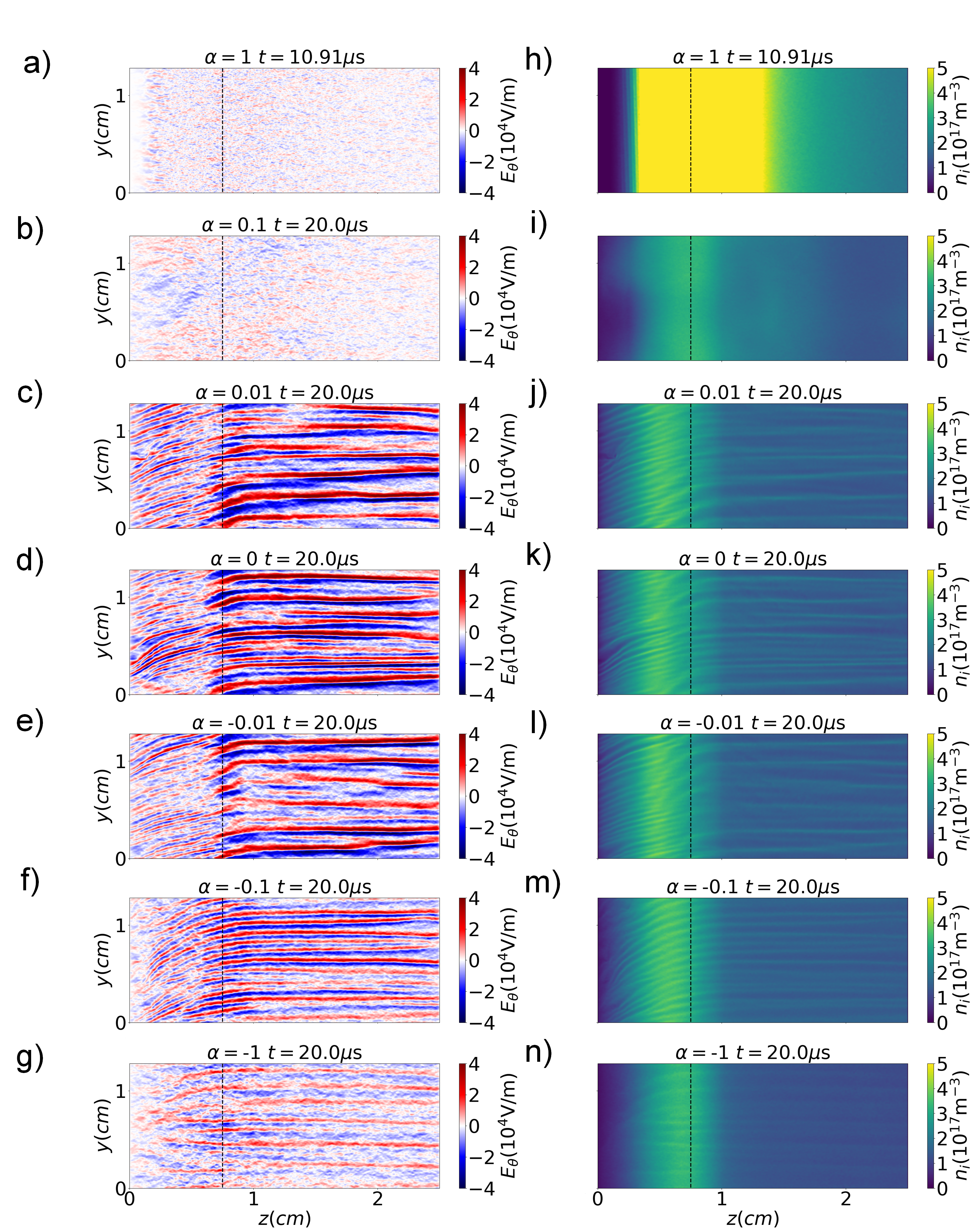}
\caption{2D maps of azimuthal electric field (a-g) and ion density (h-n), respectively correspond to $\alpha$ = +1, +0.1, +0.01, 0, -0.01, -0.1, -1 from top to bottom, the black dashed line indicates
the position of maximum magnetic field. }
\label{fig:AZ_Et_ni}
\end{figure}

The change of ion density with simulation time in the simulation domain is shown in the Fig.~\ref{fig:AZ_ni_t} \textcolor{blue}{a)}. 
Except for $\alpha$ = 1, the ion number density peaks first and then stabilizes around $2\times 10^{17}\mathrm{m^{-3}}$ in all cases.
In case $\alpha$ = 1, the ion number density is always rising, because it is almost impossible to establish an effective accelerating electric field, as shown in Fig.~\ref{fig:AZ_Et_ni} \textcolor{blue}{b)}.
Thus it is only calculated up to 10.9 $\mu$s at $\alpha$ = 1.
The steady-state ion density at $\alpha$ = -1 is slightly lower than other cases, and the axial peak electric field intensity is highest.
Additionally, the axial location of the maximum electric field strength shifts outward when $\alpha$ = -1, +0.1. 

The fluctuations of the azimuthal electric field and the ion density is closely related to the azimuthal instability.
Fig.~\ref{fig:AZ_Et_ni} shows the 2D maps of azimuthal electric field and ion density at 20 $\mathrm{\mu}$s (except the case $\alpha$ = 1 at 10.9 $\mathrm{\mu}$s).
Regardless of direction of the ion rotational flow, the amplitude of azimuthal electric field weakens as the velocity increases, as shown in Fig.~\ref{fig:AZ_Et_ni} \textcolor{blue}{a) - g)}.
It is observed that the fluctuations of the azimuthal electric field display a short wavelength structure in the positive gradient region of the magnetic field,
and then evolve into a mixed structure of long wavelength and short wavelength in the negative gradient region.
We then roughly estimate the wavelengths at the axial positions of $z/L_z$ = 0.2 and $z/L_z$ = 0.6 respectively.
When $\alpha$ = 0, $\lambda_{z/L_z=0.6} \approx 1~\mathrm{mm}$, $\lambda_{z/L_z=0.2} \approx 0.44~\mathrm{mm}$; and when $\alpha$ = $\pm$ 0.01,
$\lambda_{z/L_z=0.6} \approx 1.6~\mathrm{mm}$, $\lambda_{z/L_z=0.2} \approx 0.6~\mathrm{mm}$.
Thus when the ion rotational flow velocity increases to $\alpha$ = $\pm$ 0.01, the wavelengths of the azimuthal electric field in both positive gradient region and positive gradient region of the magnetic field become longer, although the overall azimuthal electric fields have very similar characteristics of fluctuations in Fig.~\ref{fig:AZ_Et_ni} \textcolor{blue}{c) - e)}.
When $\alpha$ = -0.1, $\lambda_{z/L_z=0.6} \approx 1~\mathrm{mm}$, $\lambda_{z/L_z=0.2} \approx 0.7~\mathrm{mm}$; when $\alpha$ = -1, $\lambda_{z/L_z=0.2} \approx \lambda_{z/L_z=0.6} \approx 1.1~\mathrm{mm}$.
As the reverse ion rotational flow velocity increases, the instability characteristics in the positive and negative gradient regions of the magnetic field become increasingly analogous.

And in the cases of the co-directional direction ion rotational flow $\alpha$ = 0.1 and 1, the fluctuations of the azimuthal electric field almost lose the characteristics of azimuthal instability. 
Because we do not consider the classical electron transport due to the collision,
the electron transport capacity is so much reduced in case $\alpha$ = 1, thus the axial electric field is very weak as illustrated Fig.\ref{fig:AZ_ni_t} \textcolor{blue}{b)}, and wide range of high density region of ions occurs as illustrated Fig.~\ref{fig:AZ_Et_ni} \textcolor{blue}{h)}.
The profiles of ion density in Fig.~\ref{fig:AZ_Et_ni} \textcolor{blue}{n) - j)} exhibit a similar trend to the azimuthal electric field.

\subsection{Effects on plasma dispersion relation}

\begin{figure}
\centering
\includegraphics[width=0.95\textwidth]{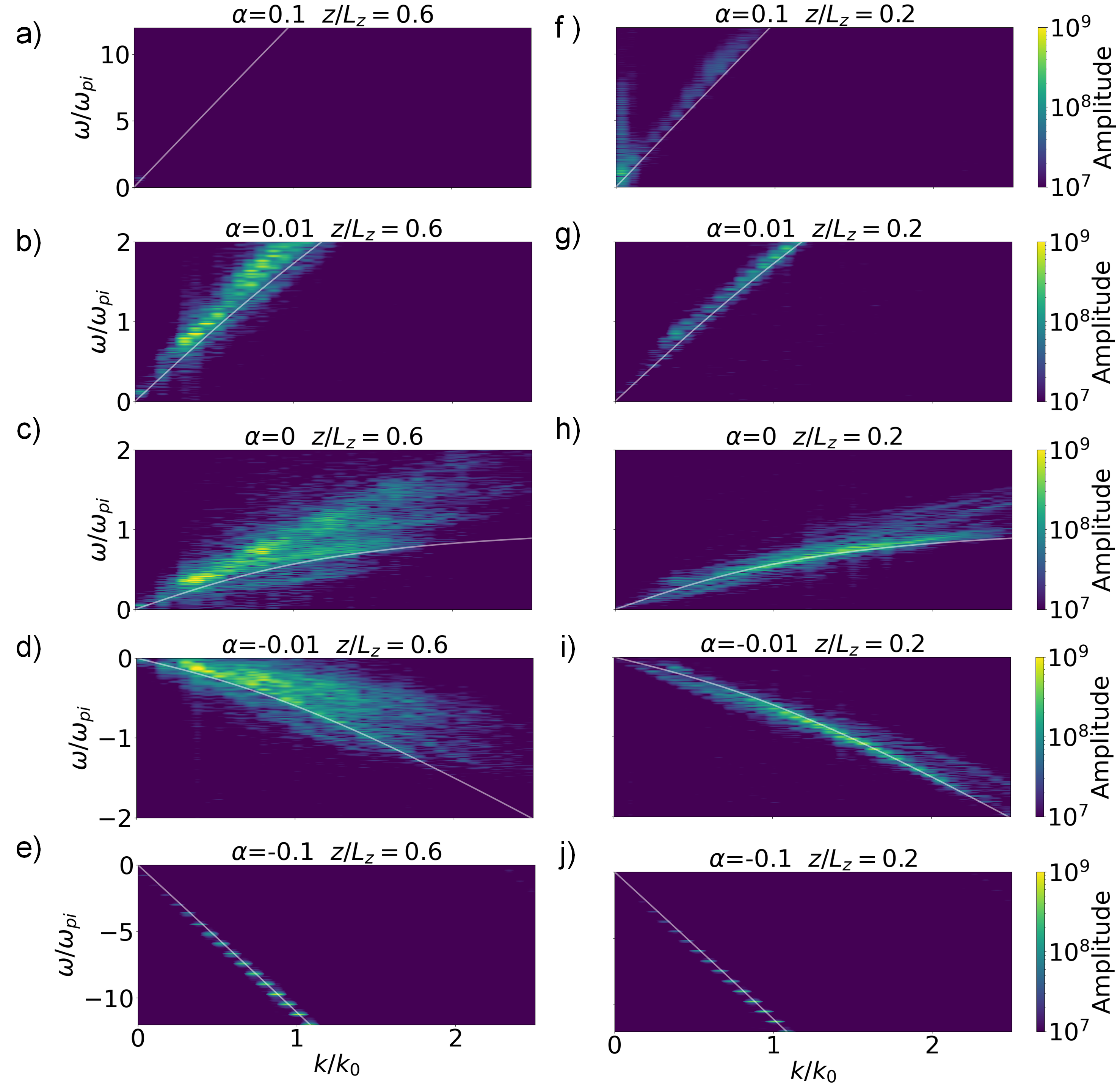}
\caption{2D maps of the FFT on the azimuthal electric field during 10-20 $\mathrm{\mu}$s in negative magnetic field gradient area ($z/L_z=0.6$) and positive magnetic field gradient area ($z/L_z=0.2$), respectively correspond to $\alpha$ = +0.1, +0.01, 0, -0.01, -0.1 from top to bottom, $k_0$ = 7035 $\mathrm{m^{-1}}$. }
\label{fig:AZ_FFT}
\end{figure}

In order to visually describe the effect of ion
rotational flow velocity on the dispersion properties of plasma, we perform the 2D fast Fourier transform (FFT) on the azimuthal electric field of 10-20 $\mathrm{\mu}$s in negative gradient region ($z/L_z$=0.6) and positive gradient region ($z/L_z$=0.2) of the magnetic field over 10-20 $\mu$s, and obtain 2D maps of dispersion relations, as illustrated Fig.~\ref{fig:AZ_FFT}. 
Here $\omega$ is normalized by $\omega_{pi}$,
$\omega_{pi}$ is the local ion plasma frequency statistically obtained from the simulation results,
and azimuthal wavenumber is normalized by $k_0 = \omega_{ce} / V_{d}$ with $B_0$ = 200 Gauss,
$E_0$ = $1\times10^4~\mathrm{V/m}$. 
The results in the condition of $\alpha$ = $\pm{1}$ are not shown, because the instability almost disappears in the condition of $\alpha$ = 1, and the dispersion relations of case $\alpha$ = -1 are similar to that of case $\alpha$ = -0.1. 
In general, the electron Bernstein wave reaches the frequency range of
ion acoustic wave 
due to electron $\mathbf{E}\times \mathbf{B}$ drift, the two resonate to form ECDI harmonics at $k=n\omega_{ce}/V_d$ (n = 1, 2, 3 $\cdots$).
In the nonlinear phase, the resonance broadening due to nonlinear effects, such as sideband instabilities induced by trapped ions\cite{2000Growth}, would result in the development of ECDI harmonics towards modified ion acoustic instability (MIAI)\cite{Lafleur_2017}.
The dispersion relation of MIAI can be approximated as follows\cite{10.1063/1.4948496,10.1063/1.5017626}
\begin{equation}\label{eq:MIAW}
\omega \approx \boldsymbol{k} \cdot \boldsymbol{V_{rot}} \pm {\frac{k c_{s}}{\sqrt{1+k^{2} \lambda_{D e}^{2}}}}
\end{equation}
The white lines in Fig.~\ref{fig:AZ_FFT} is the theoretical dispersion relation calculated using Eq.~\ref{eq:MIAW}.

Clear bright spots representing the 1st ECDI in Fig.~\ref{fig:AZ_FFT} \textcolor{blue}{b) - d)} and the resonance broadening in Fig.~\ref{fig:AZ_FFT} can be seen.
When $\alpha$ = 0, $\pm$0.01,
the dispersion relations at $z/L_z$=0.2 are more closely aligned with MIAI comparing to $z/L_z$=0.6.
The difference in the characteristics of instabilities between the positive and negative magnetic gradient regions, may be due to the coupling of several gradient drift instabilities such as magnetic field gradient, plasma density gradient, and electric field gradient.
In case $\alpha$= 0.1, higher amplitude at lower wavenumbers means that long wavelength modes are excited.
In case $\alpha$= -0.1, the instability characteristic displays a wholly distinct discrete characterisation, 
and the dispersion relation displays a high degree of similarity at the location $z/L_z$=0.2 and $z/L_z$=0.6, as illustrated Fig.~\ref{fig:AZ_FFT} \textcolor{blue}{e), j)}.
The primary distinction between the co-directional ion
rotational flow and the opposite is the direction of the waves, which depends upon the phase velocity of the waves, as follows
\begin{equation}\label{eq:phase velocity}
{v}_{phase} = V_{rot} \pm c_s 
\end{equation}
Lafleur et al.\cite{Lafleur_2017} had proven the two phase velocities ($V_{rot} + c_s$ and $V_{rot} - c_s$) correspond to a single unstable mode.
Here $T_e \approx$ 30.5 eV at steady state, and $c_s \approx 4.74 \times 10^3 ~\mathrm{m/s}$, $|V_{rot,\alpha}| \approx 8.32\times 10^5 \alpha ~\mathrm{m/s}$, according to the simulation results.
Given that $V_{rot} \gg c_s$ in the condition of $|\alpha|$ = $\pm$0.1 and $\pm$1, $v_{phase}\approx V_{rot}$.

\subsection{Effects on axial electron transport}

\begin{figure}[htbp]
    \centering
    \begin{minipage}[h]{0.35\textwidth}
    \makebox[1em]{a)}
    \centerline{\includegraphics[scale=0.35]{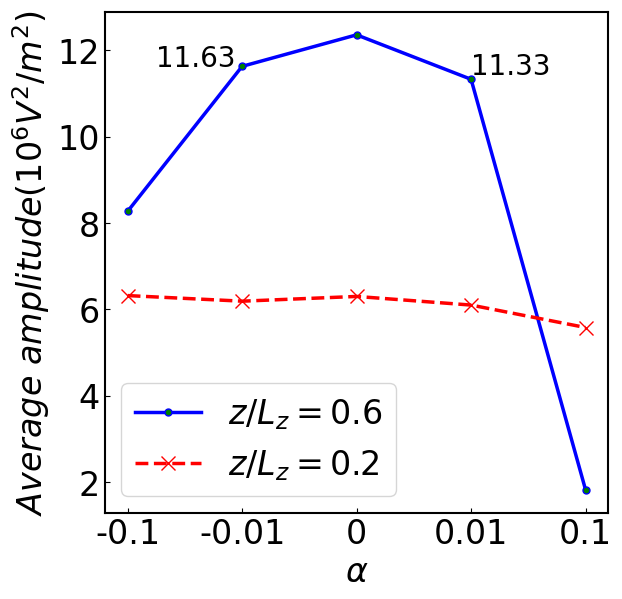}}
    \makebox[1em]{c)}
    \centerline{\includegraphics[scale=0.35]{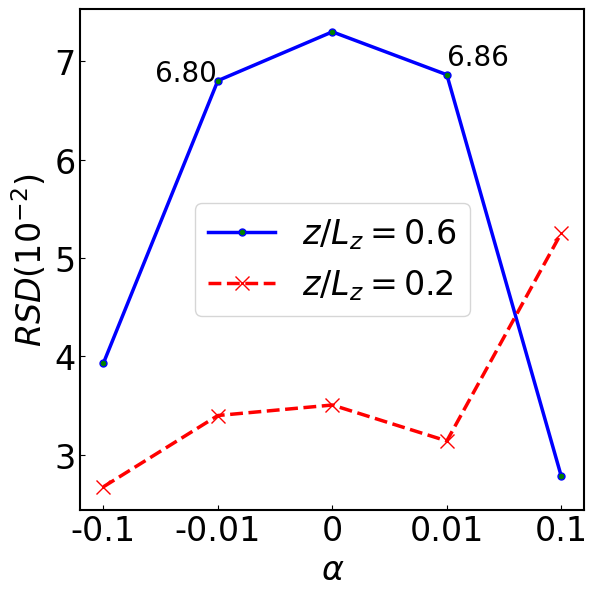}}
    \end{minipage}
    \hspace{17pt}
    \centering
    \begin{minipage}[h]{0.35\textwidth}
    \makebox[1em]{b)}
    \centerline{\includegraphics[scale=0.35]{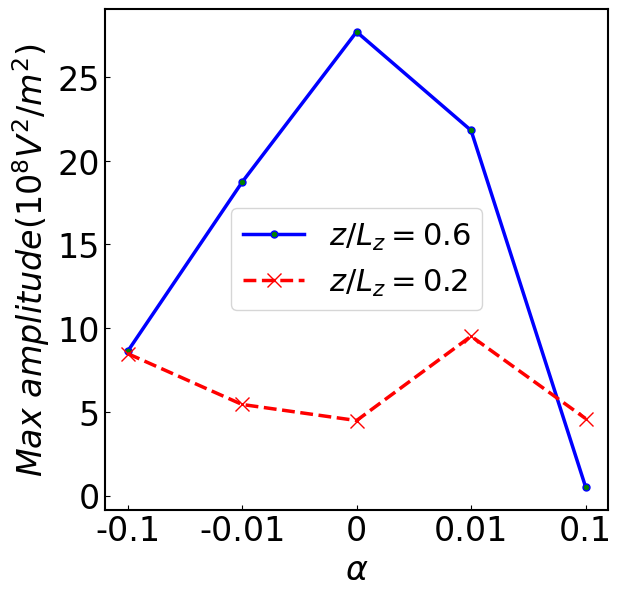}}
    \makebox[1em]{d)}
    \centerline{\includegraphics[scale=0.35]{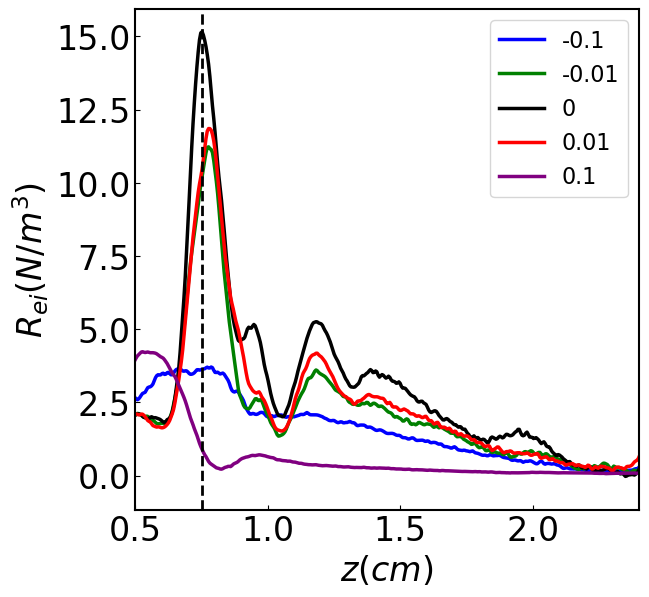}}
    \end{minipage}
\caption{a) Average amplitudes and b) maximum amplitudes of FFT on the azimuthal electric field, c) mean relative standard deviation of azimuthal electron density fluctuations, d) axial profiles of mean azimuthal effective electron-ion friction over 10-20 $\mathrm{\mu}$s, the black dashed line in d) indicates
the position of maximum magnetic field.}
\label{fig:AZ_Amp}
\end{figure}

Furthermore, the FFT results are used to calculate the average amplitude, 
which represents the average azimuthal instability intensity, as illustrated 
in Fig.~\ref{fig:AZ_Amp} \textcolor{blue}{a)}. 
The influence of ion rotational flow on the overall instability is minimal at
$z/L_z=0.2$, 
whereas at $z/L_z=0.6$, the overall instability intensity decreases with the increasing velocity of the rotational flow.
The maximum amplitude of FFT at $z/L_z$=0.6 representing the 1st ECDI intensity, the 1st ECDI intensity in case $\alpha$ = -0.01 is less pronounced than that of $\alpha$ = 0.01, as illustrated in Fig.~\ref{fig:AZ_Amp} \textcolor{blue}{b)}. 
Given that the fluctuations in azimuthal electric field and electron density are both responsible for anomalous electron transport, we plot the mean relative standard deviation of the azimuthal electron density fluctuations over 10-20 $\mathrm{\mu s}$, as illustrated in Fig.~\ref{fig:AZ_Amp} \textcolor{blue}{c)}.
As demonstrated in Fig.~\ref{fig:AZ_Amp} \textcolor{blue}{a), b), c)}, the amplitude of the fluctuations in the azimuthal electric field and electron density decreases as the ion rotational velocity increases, indicating the decrease of the azimuthal instability intensity.
Furthermore, the amplitude of the co-directional rotational flow is lower than that of the reverse for cases $| \alpha| = \pm{0.1}$, while the amplitude of the co-directional rotational flow is close to that of the reverse for cases for cases $| \alpha| = \pm{0.01}$.

Combined effects of the azimuthal electric field and electron density fluctuations, the azimuthal effective electron-ion friction is obtained from the simulation results according to the following equation\cite{10.1063/1.4948496}
\begin{equation}\label{eq:Rei}
R_{ei} =q\left\langle\delta n_{e} \delta E_{\theta}\right\rangle 
\end{equation}
As shown in Fig.~\ref{fig:AZ_Amp} \textcolor{blue}{d)}, the rise in azimuthal electron-ion friction resulting from the azimuthal instability is less pronounced in the condition of $\alpha$=-0.01 comparing to $\alpha$=0.01. 
And the friction distribution of case $\alpha$ = -0.1 corroborates that the larger reverse ion
rotational flow would lead to similar azimuthal instability characteristics of the positive and negative magnetic gradient regions. 

\begin{figure}[htbp]
    \centering
    \begin{minipage}[h]{0.5\textwidth}
    \centerline{\includegraphics[scale=0.35]{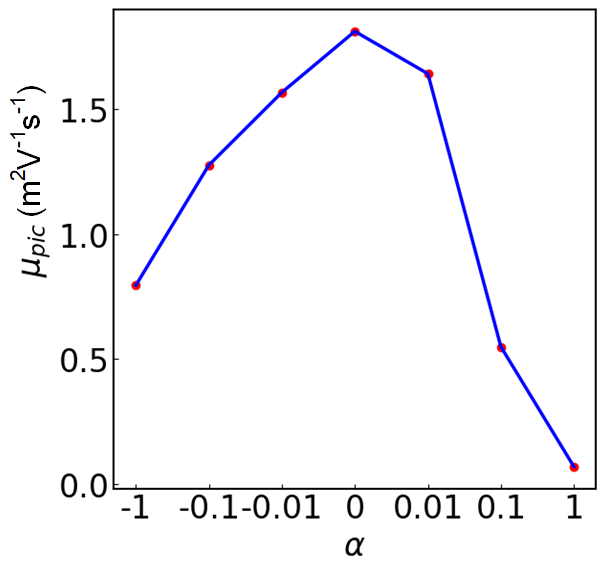}}
    \end{minipage}
\caption{ Electron mobility averaged in time and space over 10-20 $\mathrm{\mu}s$ under different ion
rotational flow velocities.}
\label{fig:AZ_mu}
\end{figure}
Finally, our research concentrate on effect of the azimuthal instability on the axial electron transport.
The electron mobility in PIC simulation is calculated using the following formula\cite{Tavant_2018}
\begin{equation}\label{eq:mu}
    \mu_{pic}=-\frac{\langle v_{e,z}\rangle }{E_{z}}
\end{equation}
$\langle v_{e,z}\rangle$ denotes the average electron velocity in the axial
direction. Spatially averaged mobility is shown in Fig.~\ref{fig:AZ_mu} \textcolor{blue}{a)}, as the ion
rotational flow velocity increases, the electron mobility decreases. 
While the electron mobility in cases $\alpha$ = 0.1 and 1 is observed to decline more significantly compared to the cases $\alpha$ = -0.1 and -1.
The electron mobility in case $\alpha$ = -0.01 is found to be less than that in case $\alpha$ = 0.01.

\section{Result analyses of radial-azimuthal simulations\label{sec:azimuthal-radial simulation results}}
\subsection{Effects on plasma profiles}

Although 2D axial-azimuthal simulations are capable of capturing changes of the instability in the axial direction, 
they are unable to account for the MTSI formed by electron-ion convection in the radial direction. 
To address this limitation, 2D azimuthal-radial simulations in different cases are conducted to investigate the impact of ion
rotational flow velocity on the coupling characteristics of the 1st ECDI and MTSI.

\begin{figure}[ht]
\centering
\includegraphics[width=0.9\textwidth]{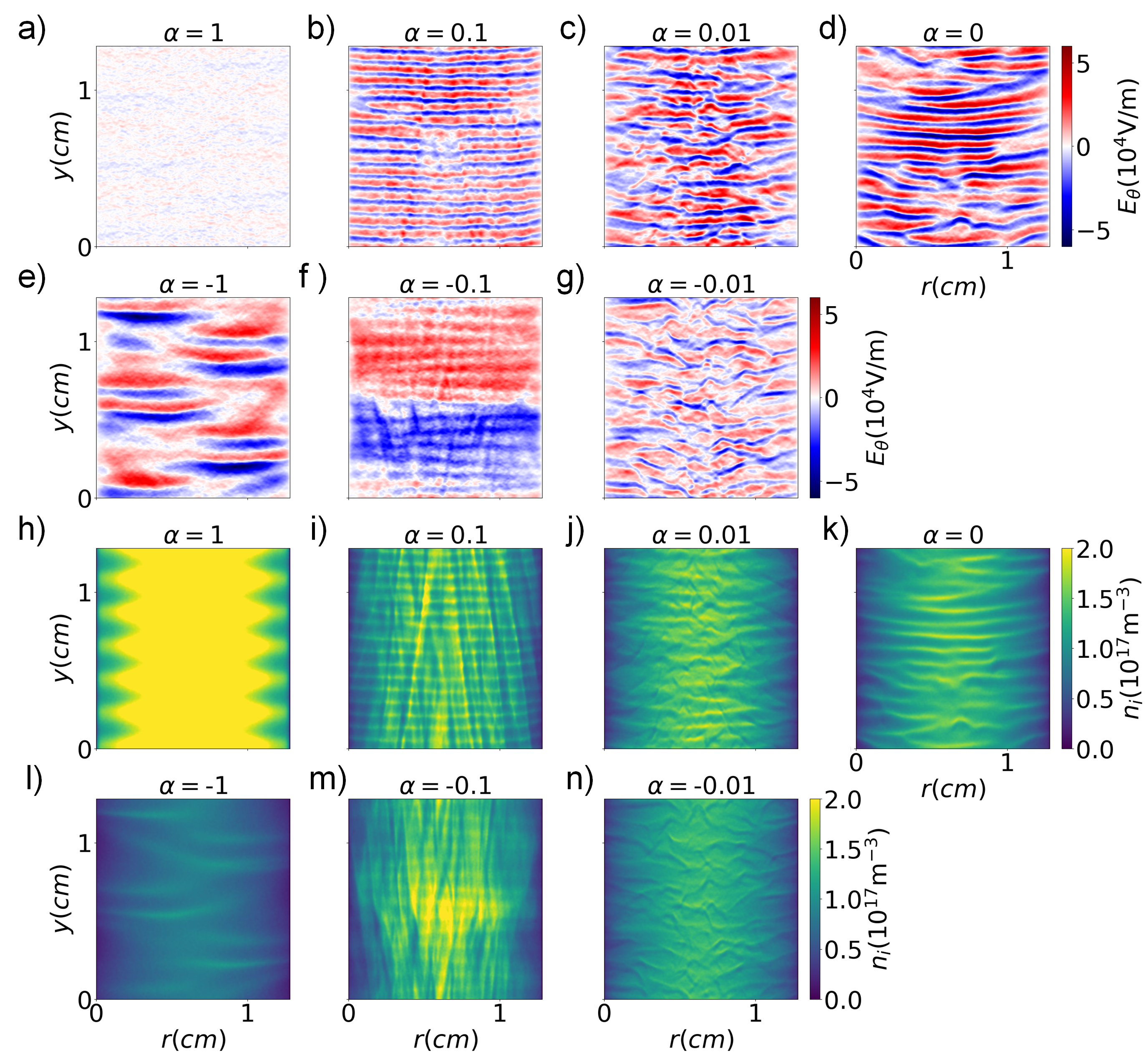}
\caption{2D maps of azimuthal electric field (a-g) and ion density (h-n) at 30 $\mathrm{\mu}$s.}
\label{fig:AR_Et}
\end{figure}

The fluctuations of azimuthal electric field and ion density illustrated in the Fig.~\ref{fig:AR_Et} at 30 $\mathrm{\mu}$s for cases $\alpha$ = 0, $\pm$0.01, demonstrate the typical characteristics associated with 1st ECDI and MTSI coupling.
As the co-directional ion
rotational flow velocity increases, the amplitude of the azimuthal electric field fluctuation decreases.
And the azimuthal electric field fluctuation gradually evolves to exhibit characteristics consistent with that of the 1st ECDI, particularly in the case of $\alpha$ = 0.1, where a regular 1st ECDI short-wavelength structure with a wavelength of 0.8 mm emerges. 
Upon increasing $\alpha$ to a value equal to the electron drifting velocity, the azimuthal instability diminishes to a point where it is no longer discernible on the Fig.~\ref{fig:AR_Et} \textcolor{blue}{a)}.
For the reverse ion
rotational flow, the 1st ECDI characteristics demonstrate a gradual weakening with the increase of the velocity,
accompanied by a gradually dominance of the radial half-wavelength characteristic associated with MTSI, particularly when $\alpha$ = -1, as illustrated in the Fig.~\ref{fig:AR_Et} \textcolor{blue}{e)}.
Additionally, a long wavelength structure spanning the entire azimuthal length with a high amplitude appears in case $\alpha$ = -0.1.

\begin{figure}[ht]
    \centering
    \begin{minipage}[h]{0.35\textwidth}
    \makebox[1em]{a)}
    \centerline{\includegraphics[scale=0.35]{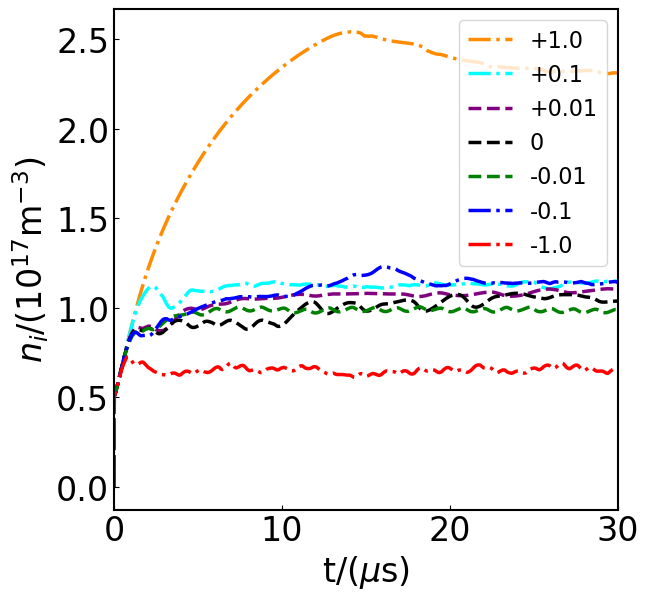}}
    \end{minipage}
    \hspace{17pt}
    \centering
    \begin{minipage}[h]{0.35\textwidth}
    \makebox[1em]{b)}
    \centerline{\includegraphics[scale=0.35]{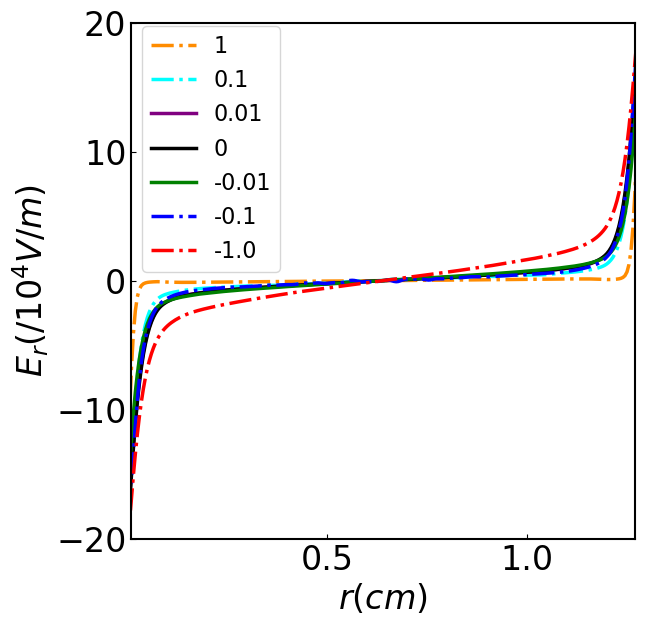}}
    \end{minipage}
\caption{a) Time profiles of ion number density for the simulation domain, b) profiles of the radial electric field averaged over 20-30 $\mu$s.}
\label{fig:AR_ni_t}
\end{figure}

In Fig.~\ref{fig:AR_ni_t} \textcolor{blue}{a)}, the ion number density is shown to stabilize around $1\times 10^{17}\mathrm{m^{-3}}$ for all cases,
except the case $\alpha$ = 1 where the ion number density increases sharply to $2.5\times 10^{17}\mathrm{m^{-3}}$, and the case $\alpha$ = -1 where the ion number density decreases significantly to around $0.6\times 10^{17}\mathrm{m^{-3}}$, which is mainly effected by the azimuthal instability. 
The ion number density has a significant impact on the plasma-wall interaction. 
In the proximity of the wall, the formation of the sheath is attributable to the higher electron density relative to that of the ions.
The increase of the ion density results in a reduction of the sheath thickness and a decline in the radial electric field intensity, as illustrated in Fig.~\ref{fig:AR_ni_t} \textcolor{blue}{b)}.

\begin{figure}[ht]
\centering
\includegraphics[width=0.9\textwidth]{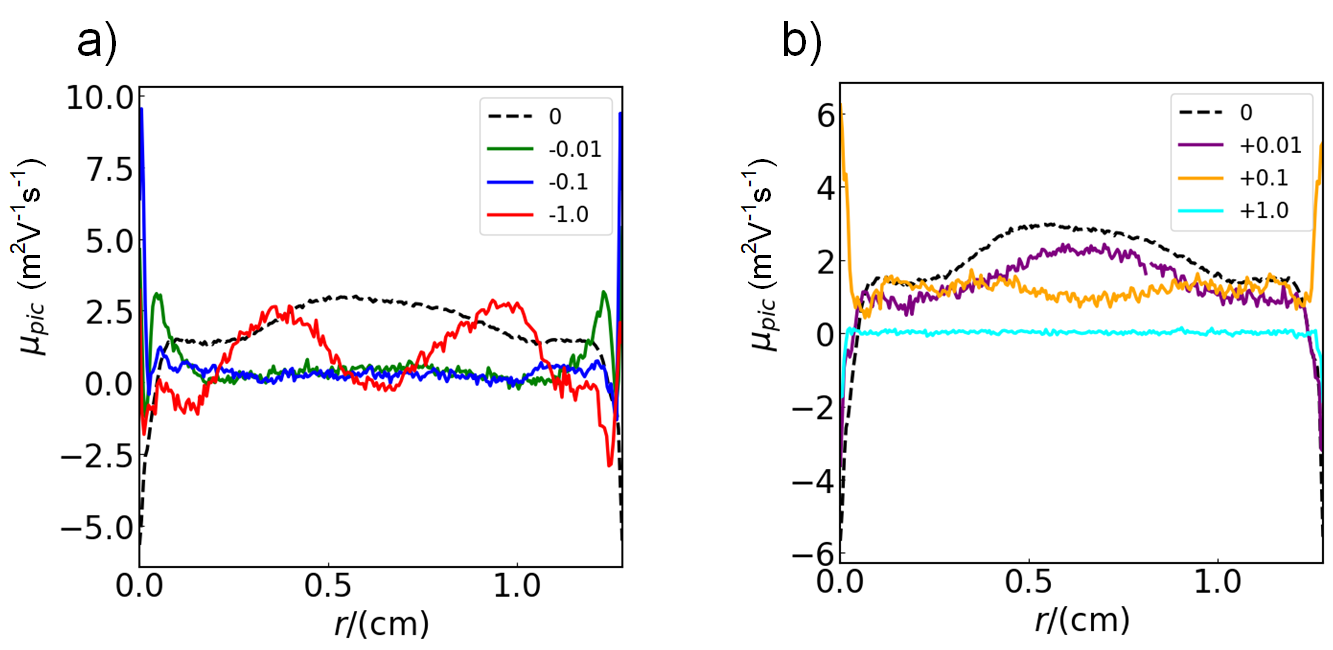}
\caption{Radial profiles of mean axial electron mobility a) in the co-directional direction, b) in the reverse direction as the electron drifting velocity over 20-30 $\mathrm{\mu}$s.}
\label{fig:AR_mu_1d}
\end{figure}

The profiles of axial electron mobility calculated from Eq.~\ref{eq:mu} along the radial direction at different ion
rotational flow velocity are illustrated in Fig.~\ref{fig:AR_mu_1d}. 
The trends of axial electron mobility with co-directional ion
rotational flow velocity are analogous to the results of the azimuthal-axial 2D simulations, where the electron mobility decreases as the co-directional ion
rotational flow velocity increases.
When $\alpha$= -0.01, -0.1, the peaks of axial electron mobility near the wall suggest that the addition of the reverse ion
rotational flow enhances the coupling of electrons to the wall sheath.
When the reverse ion
rotational flow velocity increases to the electron drifting velocity, 
the axial electron mobility fluctuates in the radial direction, 
with peaks occurring at about 0.25 and 0.75 channel width, 
which might be the result of the MTSI-dominated azimuthal instability.

%\section{Deviation from the azimuthal magnetic field theory}

\subsection{Effects on plasma dispersion relation\label{sec:n}}

\begin{figure}[ht]
\centering
\includegraphics[width=0.95\textwidth]{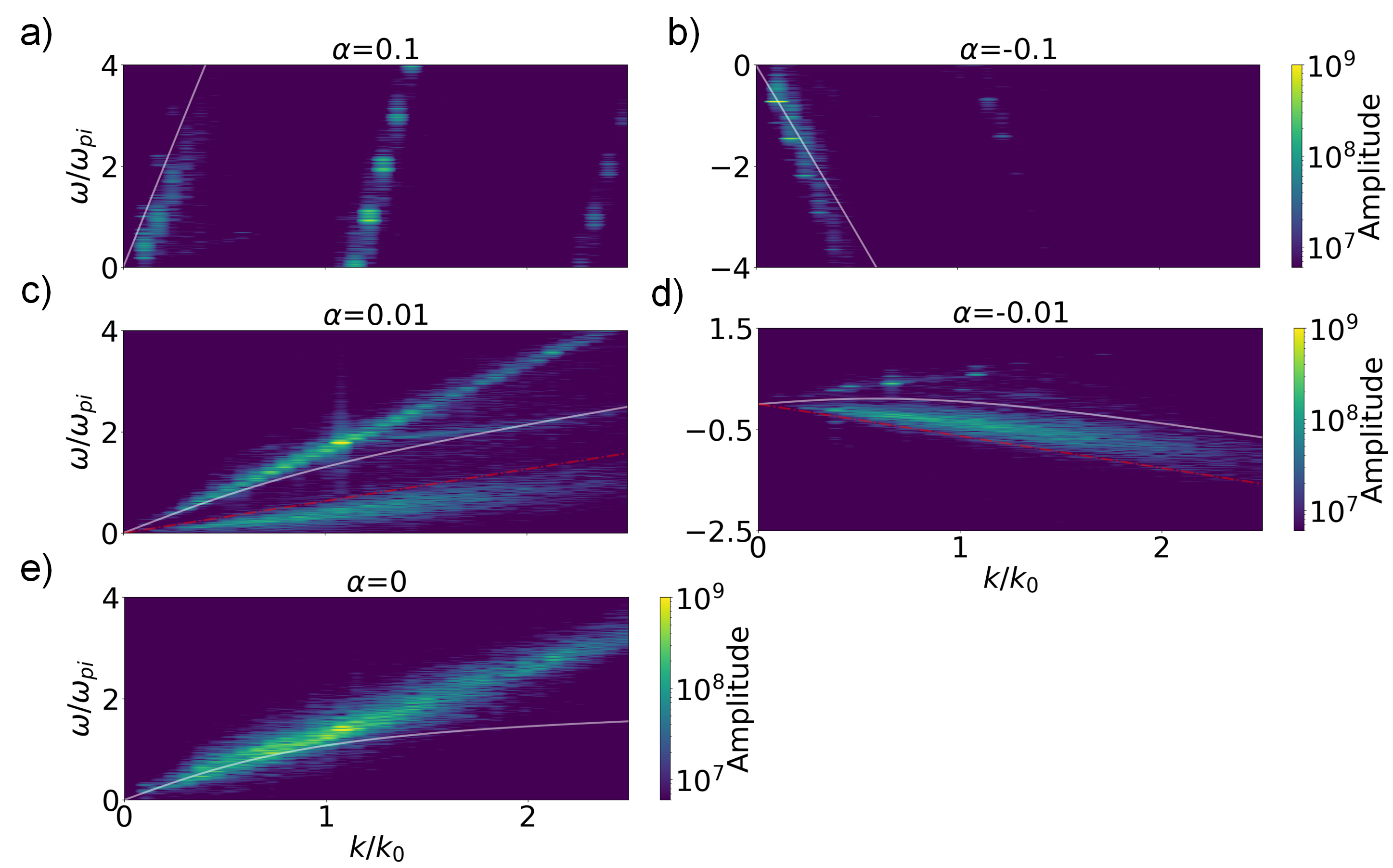}
\caption{2D maps of FFT on the azimuthal electric field during 20-30 $\mathrm{\mu}$s in the middle radial position.}
\label{fig:AR_FFT}
\end{figure}
Similarly, we performed 2D FFT on the fluctuations of azimuthal electric field over 20-30 $\mu$s in the middle radial position.
In comparison with the results obtained from 2D axial-azimuthal simulations, the images are not very close to MIAI as $k$ increases,  
which suggests that the evolution to MTAI may be primarily associated with the axial motion of ions.
In cases $\alpha = \pm0.01$, two bright bands are observed, 
one with $v_{phase} = V_{rot}$, is shown the red lines in Fig.~\ref{fig:AR_FFT} \textcolor{blue}{c), d)}, the other contains the 1st ECDI with $v_{phase} = V_{rot} + c_s$.
This phenomenon does not occur in Fig.~\ref{fig:AZ_FFT} \textcolor{blue}{b), d)} of 2D azimuthal-axial simulations, which means that the dispersion relation is further complicated by the coupling of azimuthal instability to the wall sheath. 
Compared to the 2D azimuthal-axial simulation results, the frequency bands of high amplitude is narrower in 2D radial-axial simulations.
The reason could be that for a position in the axial direction, many electrons with different drifting velocities would gather, due to the gradient of the magnetic and electric fields in 2D azimuthal-axial simulations.
Comparing the co-directional and reverse ion rotational flow, the increase of the reverse velocity enhances the long wavelength oscillation of the azimuthal electric field and suppresses the short wavelength oscillation as shown in Fig.~\ref{fig:AR_FFT} \textcolor{blue}{b), d)}, 
which is contrary to the variation law of the co-directional shown in Fig.~\ref{fig:AR_FFT} \textcolor{blue}{a), c)}.

\subsection{Effects on instability modes}

\begin{figure}[ht]
\centering
\includegraphics[width=0.95\textwidth]{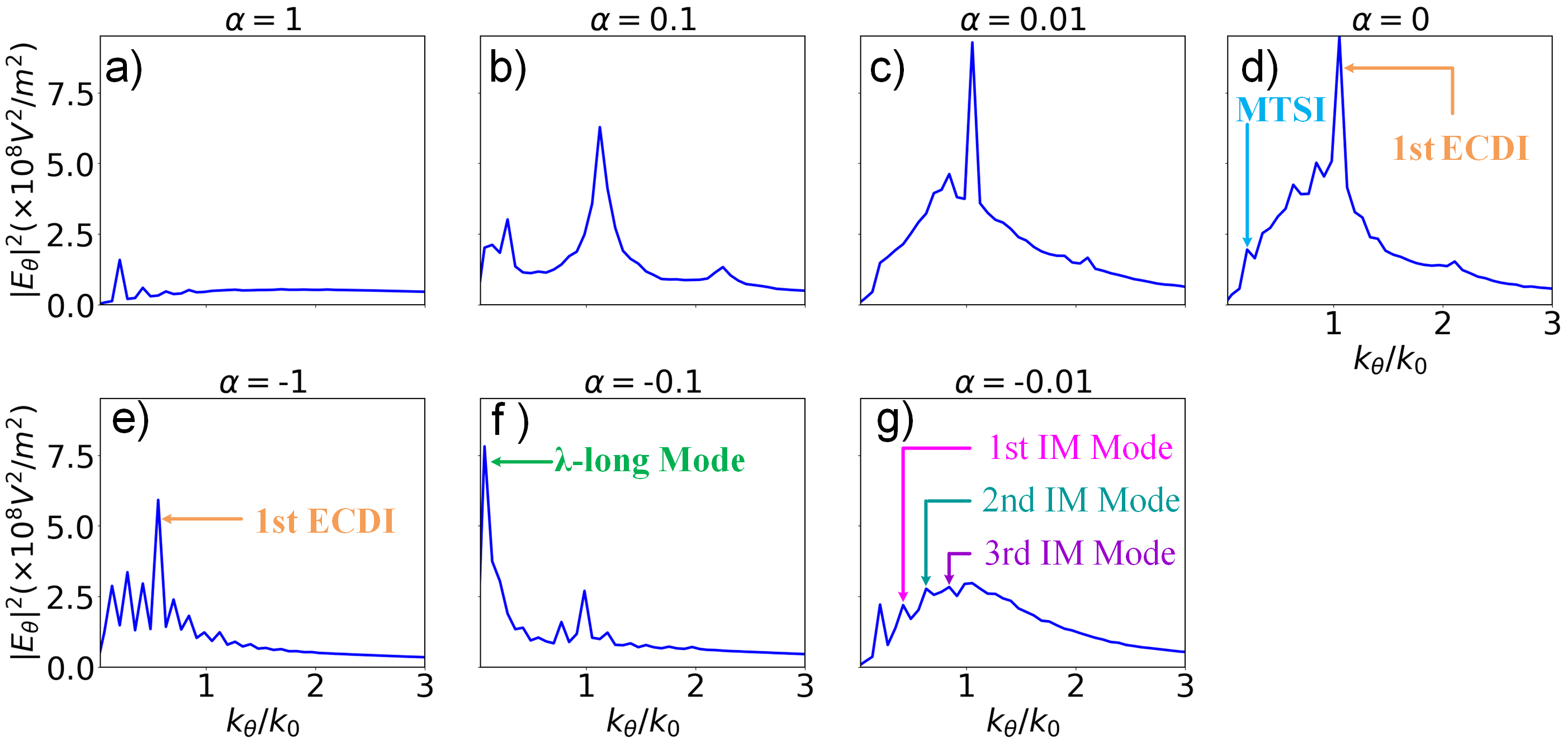}
\caption{Cumulative amplitudes of spatially FFT within the first quadrant in the azimuthal direction, the wavenumber $k_{\theta}/k_0$ is taken within 0-3.}
\label{fig:EEt_k}
\end{figure}
The above is the 2D FFT of the steady-state azimuthal electric field from the time domain to the frequency domain, we then perform the 2D FFT in space of the azimuthal electric field to observe the characteristics of the azimuthal instability at each moment.
we performed 2D map of the spatially FFT on the radial wavenumber $k_r$ and the azimuthal wavenumber $k_{\theta}$ at each moment.
Subsequently, the amplitudes of 2D spatially FFT on the azimuthal electric field within the first quadrant are aggregated in the azimuthal direction,
then we obtain the mean amplitudes of azimuthal instabilities
over 20-30 $\mu$s as illustrated in Fig.~\ref{fig:EEt_k}.

In general, the nonlinear saturation strength of azimuthal instabilities decreases with the increase of ion rotational flow velocity in Fig.~\ref{fig:EEt_k}, which is the same as the 2D azimuthal-axial simulation results.
The oscillation patterns of azimuthal instabilities exhibit a high degree of similarity of cases $\alpha$ = 0, 0.01.
For the reverse ion rotational flow, the situation is quite different.
A $\lambda$ - long mode is observed in case $\alpha$ = -0.1, which is also found in the condition of radial magnetic field gradient\cite{10.1063/5.0138223} and a certain electric field\cite{10.1063/5.0176581}.
There are three obvious intermediate modes between the 1st ECDI and MTSI, which also have been significantly affected by the ion rotational flow velocity.
In this paper, we name the three intermediate modes in order of wavenumbers from low to high as the first intermediate mode ($\mathrm{1st~IM~mode}$), the second intermediate mode ($\mathrm{2nd~IM~mode}$), the third intermediate mode ($\mathrm{3rd~IM~mode}$), as shown in Fig.~\ref{fig:EEt_k} \textcolor{blue}{g)}.
At $\alpha$ = -1, there are only three peaks before 1st ECDI, the MTSI and intermediate modes cannot be clearly captured, as shown in Fig.~\ref{fig:EEt_k} \textcolor{blue}{e)}.
At $\alpha$ = 1, the intensity of azimuthal instabilities is small.
In view of the above conditions, we only analyze the instability modes for every case which can be determined by the wavenumber, while the MTSI of case $\alpha$ = -1, the 1st ECDI and MTSI of case $\alpha$ = 1  can not.
According to Fig.~\ref{fig:EEt_k}, we obtain the wavenumbers of the 1st ECDI, MTSI, intermediate modes, $\lambda$ - long mode for each case, as shown in Tab.~\ref{tab:k}.
Tab.~\ref{tab:k} verifies the wavenumber shift of the 1st ECDI and MTSI predicted by the theory in Fig.~\ref{fig:wavenumber}.
Due to the limitation of the finite radial size, the minimum sampling interval is 0.07, so the wavenumber shift is not obvious in cases $\alpha$ = 0, $\pm$0.01.
\begin{table}[]
	\centering
        \caption{Normalized wavenumbers $k_{\theta}/k_0$ of azimuthal instability modes for each case based on the simulation results.}
	\begin{tabular}{ccccccc}
		\toprule
$\alpha$ & -1    & -0.1  & -0.01, 0, and 0.01 & 0.1 \\ \midrule
1st ECDI  & 0.56  & 0.98  & 1.05 & 1.12\\
MTSI  & / & 0.14 & 0.21 & 0.28 \\
		1st IM mode & /  & 0.21 & 0.28 & 0.35 \\
  2nd IM mode & /  & 0.56   & 0.63  & 0.7 \\
  3rd IM mode & /  & 0.77   & 0.84  & 0.91  \\
  $\lambda$ - long mode & / & 0.07  & / & / \\
  \bottomrule
	\end{tabular}
	\label{tab:k}
\end{table}

\begin{figure}[ht]
\centering
\includegraphics[width=0.9\textwidth]{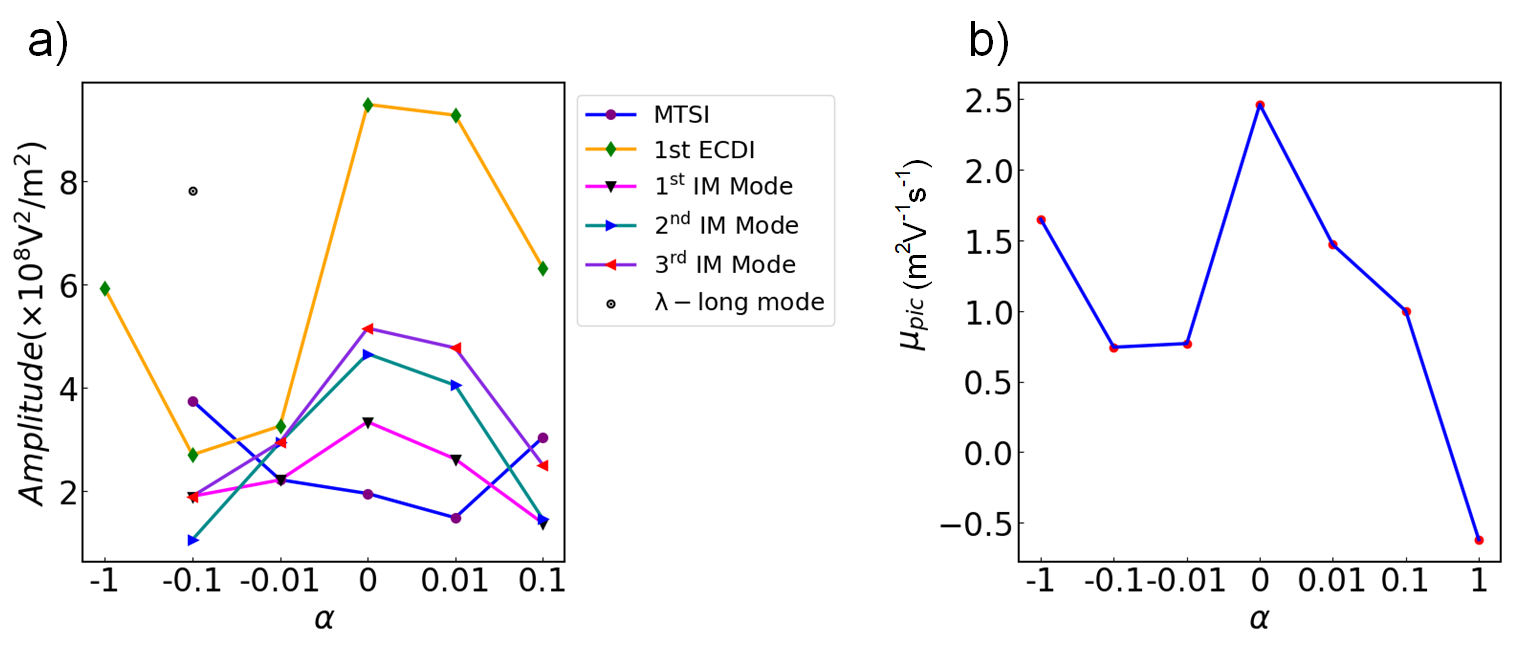}
\caption{a) Average amplitudes of the 1st ECDI, MTSI, intermediate modes, $\lambda$-long mode, b) average electron axial mobility of different ion rotational flow velocities in the saturation phase over 20-30 $\mu$s.}
\label{fig:AR_mu}
\end{figure}

We then pay more attention to the influence of various instabilities on the axial electron transport in the nonlinear saturation stage, as shown in Fig.~\ref{fig:AR_mu}. 
With the increase of rotational velocity for $\alpha$ = 0, $\pm$0.01, $\pm$0.1, the intensity of the 1st ECDI and intermediate modes decreases, as shown in Fig.~\ref{fig:AR_mu} \textcolor{blue}{a)}.
For the co-directional and reverse ion rotational flow at $|\alpha|$ = 0.01, 
although the strength of MTSI for the reverse is slightly higher, the strength of the 1st ECDI is obviously weaker, so the axial electron mobility of the reverse is lower, as shown in Fig.~\ref{fig:AR_mu} \textcolor{blue}{b)}.
And the axial electron mobility does not seem to be correlated with the amplitude of the $\lambda$ - long mode in case $\alpha$ = -0.1.
In general, the axial electron mobility is mainly determined by the 1st ECDI for the co-directional ion rotational flow, and is affected by both MTSI and 
the 1st ECDI due to the increase of MTSI relative to the 1st ECDI for the reverse.

% For the radial non-uniform distribution of $\overline{v}_z$,
% a more careful study is needed but will be left as a future work,
% which may involve the numerical drawbacks of
% using the RZ dimension or the way of setting the macro-particle weights,
% which are beyond the scope of this paper.
\subsection{Comparison between theory and simulation}

\begin{figure}[htbp]
    \centering
    \begin{minipage}[h]{0.35\textwidth}
    \makebox[1em]{a)}
    \centerline{\includegraphics[scale=0.32]{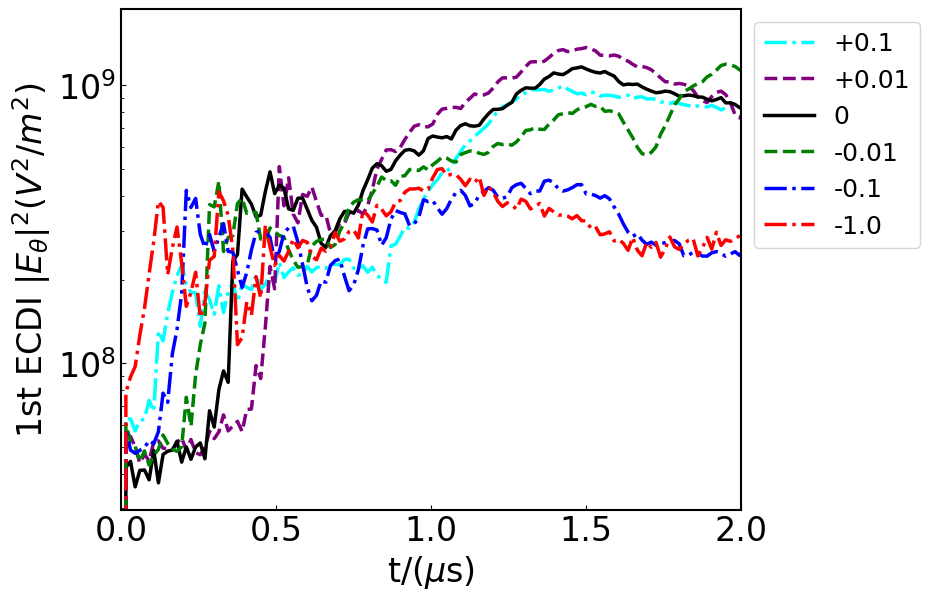}}
    \end{minipage}
    \hspace{50pt}
    \centering
    \begin{minipage}[h]{0.35\textwidth}
    \makebox[1em]{b)}
    \centerline{\includegraphics[scale=0.32]{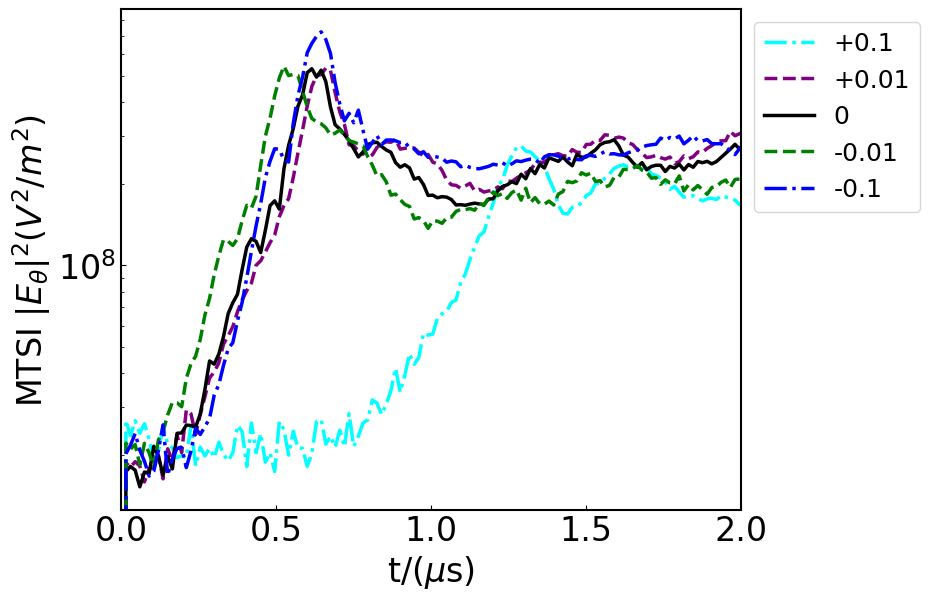}}
    \end{minipage}
\caption{a) Amplitudes profiles of the 1st ECDI, b) amplitudes profiles of MTSI over 0-2 $\mathrm{\mu}$s.}
\label{fig:AR_EET0-2}
\end{figure}

Fig.~\ref{fig:AR_EET0-2} illustrates the temporal evolution of the 1st ECDI and MTSI over 0-2 $\mu$s.
For the reverse ion rotational flow, it is found that the earlier the linear phases of 1st ECDI are excited with the greater velocity between ions and electrons.
We find that the slope of the linear growth phase for the 1st ECDI does not appear to undergo a significant change for the opposite ion
rotational flow.
However, an increase in ion
rotational flow velocity from $\alpha$ = -0.1 to -1 is observed to result in a corresponding increase in growth rate, from 0.5 to 1.078
as the theoretical predictions in Sec.~\ref{sec:theory}.
Therefore, we have obtained the theoretical and simulated linear phase growth rates below for comparison.

As indicated in the Ref.~\cite{Villafana_2021}, the amplitudes of MTSI and 1st ECDI in the linear growth stage exhibits an exponential relationship with their respective growth rates $\mathrm{A = C exp}(2\gamma t)$
(the constant C is devoid of any physical significance).  
Based on Fig.~\ref{fig:AR_EET0-2}, the time corresponding to the linear growth stages under different ion rotational flow velocities for 1st ECDI and MTSI is selected respectively, and the least squares method is employed to fit the growth rates.
The theoretical growth rates are obtained through the following process: the mean electron density and mean electron temperature over the linear growth period are obtained as shown in Tab.~\ref{tab:ne_te} and entered into iterative Eq.~\ref{eq:iterative equation} for solution. 
In the initial phase of instability development (0-1 $\mu$s), ion trapping has not yet occurred, and the ion azimuthal velocity is equal to the initial ion rotational flow velocity.

\begin{table}[]
	\centering
        \caption{Mean electron temperature and electron density in the linear growth phase under different ion
rotational flow velocity.}
	\begin{tabular}{ccccccc}
		\toprule
		$\alpha$       & -1    & -0.1  & -0.01 & 0     & 0.01  & 0.1  \\ \midrule
MTSI $\mathrm{n_e}$($\times10^{16}\mathrm{m^{-3}}$)  & / & 6.27  & 6.58  & 6.93  & 6.95  & 9    \\
MTSI $\mathrm{T_e}$(eV) & / & 10.76 & 12.65 & 12.83 & 12.26  & 19   \\
		1st ECDI $\mathrm{n_e}$($\times10^{16}\mathrm{m^{-3}}$) & 5.15  & 5.5   & 5.79  & 6.21  & 6.73  & 5.3 \\
		1st ECDI $\mathrm{T_e}$(eV) & 10.08 & 9.76  & 10.11 & 9.85  & 10.73 & 9.74 \\ \bottomrule
	\end{tabular}
	\label{tab:ne_te}
\end{table}

\begin{figure}[ht]
\centering
\includegraphics[width=0.35\textwidth]{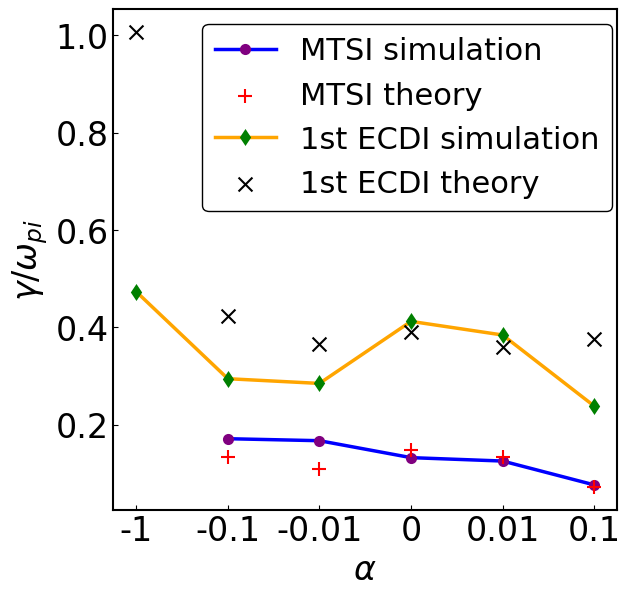}
\caption{Comparison of theoretical and simulated growth rates in the linear phase.}
\label{fig:growth_rate_comparison}
\end{figure}

The theoretical and simulated growth rates are shown in Fig.~\ref{fig:growth_rate_comparison}, here the growth rates of MTSI and 1st ECDI are normalized with $\omega_{pi}$, taking $n_0=1.2\times10^{17}~\mathrm{m^{-3}}$ for MTSI and $n_0=0.817\times10^{17}~\mathrm{m^{-3}}$ for the 1st ECDI.
For the 1st ECDI, the theoretical and simulated growth rates are close at $\alpha$ = 0, 0.01, but the theoretical growth rates are significantly greater at $\alpha$ = 0.1, -0.01, -0.1, -1.
On the contrary, simulated growth rates of MTSI are greater than the theoretical at $\alpha$ = -0.01, -0.1, -1.

\begin{figure}[ht]
\centering
\includegraphics[width=0.95\textwidth]{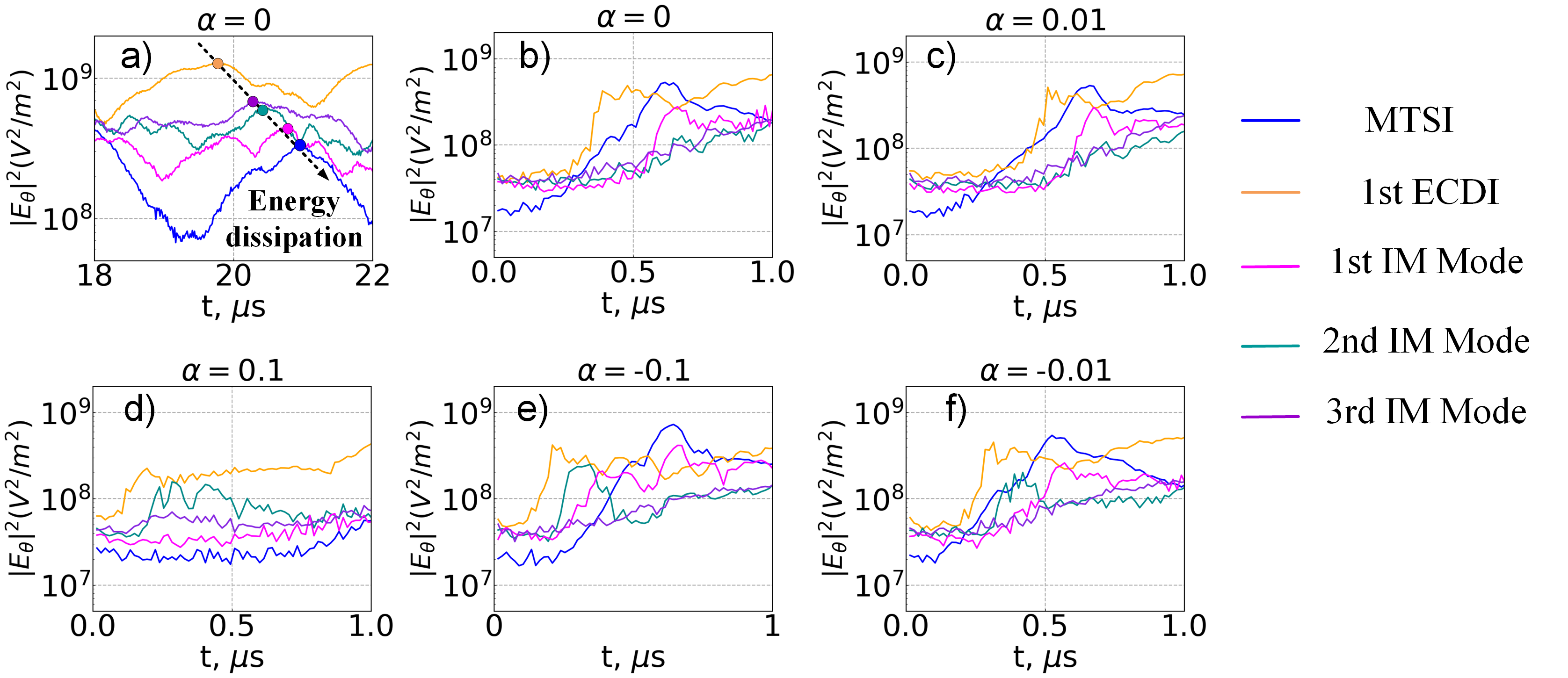}
\caption{ Amplitude profiles of MTSI, the 1st ECDI, $\lambda$ - long mode, and intermediate modes in 0-1 $\mu$s. }
\label{fig:AR:EEt0-1}
\end{figure}

The formation and development of the unstable modes are effected by wave-wave resonance and energy inverse cascade effect in plasma turbulence.
The wave energy is transferred from short wavelength waves to long wavelength waves, and the intermediate wave-wave resonance modes establish the mechanism of energy dissipation from the 1st ECDI to MTSI, as shown in Fig.\ref{fig:AR:EEt0-1} \textcolor{blue}{a)}.
The evolution of intermediate modes with time shows the characteristics of 
the 1st ECDI and MTSI coupling.
For cases $\alpha$ = 0, 0.01, the evolution trend of intermediate modes in the linear stage is basically the same, as shown in Fig.\ref{fig:AR:EEt0-1} \textcolor{blue}{b), c)}.
For cases $\alpha$ = 0.1, -0.01, -0.1, the intensity of intermediate modes (especially the 2nd IM mode) increases significantly, as shown in Fig.\ref{fig:AR:EEt0-1} \textcolor{blue}{d), e), f)}.
The energy of the 1st ECDI is dissipated to intermediate modes in the linear growth stage due to the resonance between the waves, thus the simulated growth rate of the 1st ECDI is less than the theoretical.
And as the reverse ion
rotational flow velocity increases, 
this energy dissipation increases,
the difference between simulated and theoretical growth rates becomes larger.
For cases $\alpha$ = -0.01, -0.1, 0.1, the simulated growth rates of MTSI are greater than the theoretical because of the energy transfer of the 1st ECDI to longer wavelength modes.
According to above ansyses, ion rotational flow increases the nonlinear effect and complicates the wave-wave interactions between the azimuthal instabilities in the linear stage.

\section{Discussions and Conclusions\label{sec:conclusion}}
Before drawing conclusions, for the ion rotational flow velocity magnitude $|\alpha|$ = 0.01, we evaluate the effects of the azimuthal instability on the electron mobility.
Fig.~\ref{fig:AZ_mu} and Fig.~\ref{fig:AR_mu} have shown the electron mobility induced by the azimuthal instability, below we calculate the rate of change of the electron mobility. The rate of change with the addition of the ion rotational flow relative to the absence is represented by $\eta = (\mu_{\alpha=0}-\mu_{\alpha})/\mu_{\alpha=0}$, which with the reverse ion rotational flow relative to the co-directional is represented by $\eta ^{'} = (\mu _{\alpha = 0.01}- \mu _{\alpha = -0.01})/\mu _{\alpha = 0.01}$.
$\eta _{\mathrm{AZ}, \alpha = -0.01} = 13.5\%$, $\eta _{\mathrm{AZ}, \alpha = 0.01} = 9.4\%$, $\eta^{'}$ = 4.3\% for 2D azimuthal-axial simulations; $\eta _{\mathrm{AR}, \alpha = -0.01} = 68.6\%$, $\eta _{\mathrm{AZ}, \alpha = 0.01} = 40.2\%$, $\eta^{'}$ = 31\% for 2D azimuthal-radial simulations. 
And $\eta_{\mathrm{AZ}}/\alpha \approx10$, $\eta_{\mathrm{AR}}/\alpha \approx40$, which means 1\% change of ion rotational flow velocity results in a 10\% change in electron mobility for azimuthal-axial simulations, and a more significant 40\% change for azimuthal-radial simulations.
Combined with the change of azimuthal instability, we believe that for the reverse ion rotational flow, although the intensity of MTSI increases slightly, the intensity of the 1st ECDI decreases more than that of the co-directional ion rotational flow, 
so the axial electron mobility is lower, which is an important mechanism for the HT to perform better under reverse rotational flow than co-directional rotational flow.

We draw the following major conclusions
based on the results obtained in this work:
\begin{itemize}
  \item For 2D azimuthal-axial simulations, the azimuthal instability weakens with the increase of the velocity for both reverse and co-directional ion rotational flow.
  When the reverse ion rotational flow velocity is large enough ($\alpha \leq -0.1$), the dispersion relations and instability characteristics in the positive and negative magnetic field gradient regions become similar. 
  The phase velocity of the dispersion relation becomes $ v_{phase}=V_{rot} + c_s$ with the addition of the ion rotational flow, thus the direction of rotational flow determines the direction of wave propagation for cases $|\alpha|$ = 0.1, 1.
  \item For 2D azimuthal-radial simulations, the reverse ion rotational flow 
affects plasma-wall interaction, resulting in the growth of the near-wall sheath, while the near-wall sheath is inhibited for the co-directional.
 Thus for the co-directional ion rotational flow, the 1st ECDI dominates the electron anomalous transport; for the reverse, near-wall sheath growth leads to MTSI growth, and the instability exhibits a more complex coupling mode of MTSI and the 1st ECDI. And a $\lambda$ - long mode occurs in case $\alpha$ = -0.1. 
  \item Theoretical and simulation results demonstrate that the wavenumbers of the 1st ECDI and MTSI are monotonically decreasing functions of the velocity of ions relative to electrons.
    \item The ion rotational flow strengthens the wave-wave interactions between unstable modes for cases $\alpha$ = 0.1, -0.01, -0.1. The energy transfers from the 1st ECDI to intermediate resonance modes and MTSI results in the simulated growth rates of 1st ECDI less than the corresponding kinetic theor.
The energy is further dissipated to MTSI, resulting in simulated growth rates of MTSI greater than the corresponding kinetic theory.
\end{itemize}

Comparing the dispersion relations of the axial-azimuthal and radial-azimuthal simulations, we can see the important influence of the axial electric field gradient, magnetic field gradient, and density gradient on the evolution of the instability to MIAI, and the important influence of the wall sheath on the coupling characteristics of the 1st ECDI and MTSI.
In the future, additional studies will need to be conducted in order to gain a deeper understanding of the coupling mechanism between the azimuthal instability and wall sheath, and the influence of nonlinear effects on instability in the presence of the ion
rotational flow.

\section*{Acknowledgment}
The authors acknowledge the support from National Natural Science Foundation of China (Grant No.5247120164). 
This research used the open-source particle-in-cell code WarpX \href{https://github.com/ECP-WarpX/WarpX}{https://github.com/ECP-WarpX/WarpX}, primarily funded by the US DOE Exascale Computing Project. Primary WarpX contributors are with LBNL, LLNL, CEA-LIDYLSLAC, DESY, CERN, and TAE Technologies. We acknowledge all WarpX contributors.

\section*{Data Availability}
The data that support the findings of this studyis available from the corresponding author upon reasonable request.

\section*{Reference}

\bibliographystyle{elsarticle-num}
\bibliography{reference}

\end{document}